\documentclass[10pt,aip,twocolumn]{revtex4-1}
\newcommand{\V}[1]{\mathbf{#1}} 
\newcommand\Alfven{Alfv\'en } 
\newcommand\Alfvenic{Alfv\'enic } 
\newcommand\tempAni{$T_{\perp p}/T_{\parallel p}$ } 
\newcommand\Bpar{$\beta_{\parallel p}$ } 
\usepackage{fullpage}
\usepackage{amssymb}
\usepackage{amsmath}
\usepackage{hyperref}
\usepackage{natbib}
\usepackage{color}
\usepackage{graphicx}
\usepackage{graphics}

\begin{document}

\title{Predicted Impacts of Proton Temperature Anisotropy on Solar Wind Turbulence}
\author{K.~G. Klein}
\email{kristopher.klein@unh.edu}
\affiliation{Space Science Center, University of New Hampshire, Durham,
NH 03824, USA}

\author{G.~G. Howes}
\affiliation{Department of Physics and Astronomy, University of Iowa, Iowa City, 
Iowa 52242, USA.}


\begin{abstract}
Particle velocity distributions measured in the weakly collisional
solar wind are frequently found to be non-Maxwellian, but how these
non-Maxwellian distributions impact the physics of plasma turbulence
in the solar wind remains unanswered. Using numerical solutions of the
linear dispersion relation for a collisionless plasma with a
bi-Maxwellian proton velocity distribution, we present a unified
framework for the four proton temperature anisotropy instabilities,
identifying the associated stable eigenmodes, highlighting the
unstable region of wavevector space, and presenting the properties of
the growing eigenfunctions.  Based on physical intuition gained from
this framework, we address how the proton temperature anisotropy
impacts the nonlinear dynamics of the \Alfvenic fluctuations
underlying the dominant cascade of energy from large to small scales
and how the fluctuations driven by proton temperature anisotropy
instabilities interact nonlinearly with each other and with the
fluctuations of the large-scale cascade. We find that the nonlinear
dynamics of the large-scale cascade is insensitive to the proton
temperature anisotropy, and that the instability-driven fluctuations
are unlikely to cause significant nonlinear evolution of either the
instability-driven fluctuations or the turbulent fluctuations of the
large-scale cascade.
\end{abstract}

\maketitle

\section{Introduction}
\label{sec:intro}
The near-Earth solar wind is a dynamic plasma environment supporting
broadband turbulent spectra of plasma and electromagnetic field
fluctuations, providing a uniquely accessible venue for the study of
the fundamental physics of astrophysical plasma turbulence. Direct
spacecraft measurements show that the particle velocity distributions
of ions and electrons in the solar wind commonly deviate from the
isotropic Maxwellian velocity distributions characteristic of a plasma
in local thermodynamic equilibrium, a result that is not unexpected
considering the weak collisionality of the solar wind plasma.  But how
the non-Maxwellian nature of the plasma particle distribution
functions impacts the physics of plasma turbulence in the solar wind
remains unanswered.  Unraveling the nature of turbulence in the solar
wind is a grand challenge problem in heliophysics because turbulence
significantly impacts the transport of energy from large-scale motions
to sufficiently small scales at which that energy is efficiently
converted to plasma heat or some other non-thermal form of particle
energization.

Turbulence in the solar wind is dominated by energy injected into the
turbulent cascade at large scales through nonlinear interactions among
plasma motions at those large scales. The turbulent cascade from those
large scales down to characteristic ion kinetic scales, denoted the
\emph{inertial range} of the turbulence, occurs without significant
dissipation and is mediated by nonlinear interactions between
incompressible \Alfvenic fluctuations with wavevectors that become
increasingly perpendicular ($k_\perp \gg k_\parallel$, where
perpendicular and parallel are defined relative to the local mean
magnetic field) with diminishing length scale. In addition, this
turbulent cascade of energy from large scales through the inertial
range also includes a small admixture of compressible fluctuations,
which appear to have similarly anisotropic wavevectors with $k_\perp
\gg k_\parallel$.\cite{Howes:2012a,Klein:2012} 
Separate from this
anisotropic cascade of energy from large scales (referred to as the
\emph{large-scale cascade} in this paper), kinetic instabilities
driven by non-Maxwellian velocity distributions may also inject energy
directly into the turbulence typically at small scales near the characteristic
ion or electron kinetic scales.

As the first step of an in-depth study of the impact of non-Maxwellian
distribution functions on plasma turbulence in the solar wind, we
focus here specifically on the effect of a bi-Maxwellian proton
temperature distribution. We investigate two open questions: (1) How
does the proton temperature anisotropy impact the nonlinear dynamics
of the \Alfvenic fluctuations underlying the large-scale cascade? (2)
What is the nature of the modes generated by kinetic instabilities
driven by the proton temperature anisotropy, and how do they
contribute to the fluctuations measured in the solar wind?

In Section~\ref{sec:physics}, we discuss the physics of the proton
temperature anisotropy and the resulting kinetic instabilities, and we
describe the numerical approach used to solve for the linear kinetic
plasma behavior. A unified framework for understanding the four proton
temperature anisotropy instabilities is presented in
Section~\ref{sec:fluid}.  The physical properties described in this
framework are employed to discuss the effect of the proton temperature
anisotropy on the large-scale turbulent cascade in
Section~\ref{sec:turb} and to examine how fluctuations generated by
the kinetic instabilities interact nonlinearly with each other and
with the turbulent fluctuations of the large-scale cascade in
Section~\ref{sec:injection}. Our findings are summarized in
Section~\ref{sec:conc}, and a brief technical note on the
identification of normal modes of the Vlasov-Maxwell system appears in
Appendix~\ref{sec:EPs}.

\section{The Physics of Proton Temperature Anisotropy}
\label{sec:physics}

Compressions or expansions of a magnetized plasma result in different
rates of change of charged particle velocities in the directions
perpendicular and parallel to the local magnetic field. Collisions act
to isotropize the velocity distributions in different directions,
ultimately driving the velocities to a Maxwellian distribution,
corresponding to local thermodynamic equilibrium with a single
isotropic temperature, a state of maximum entropy from which no free
energy can be extracted. But, in the solar wind, collisions are often
insufficient to yield isotropy between parallel and perpendicular
temperatures, representing a potential source of free energy in the
particle velocity distributions. If the plasma motions lead to
temperatures that are sufficiently anisotropic, then kinetic
temperature anisotropy instabilities can tap this source of free
energy to drive electromagnetic fluctuations in the plasma that
ultimately serve to reduce the temperature anisotropy.  

Note that other types of non-Maxwellian velocity distributions, such
as the presence of a beam component or a relative drift between plasma
species, also contain free energy that may drive kinetic
instabilities. But, to build a foundation upon which to understand the
effect of non-Maxwellian velocity distributions on plasma turbulence,
we begin with the idealized case of a bi-Maxwellian proton temperature
distribution, in which the free energy content is characterized by the
single parameter $T_{\perp p}/T_{\parallel p}$.

Spacecraft measurements in the near-Earth solar wind demonstrate a
wide spread of values of the proton temperature anisotropy $T_{\perp
  p}/T_{\parallel p}$, often filling the entire range between the marginal
stability boundaries of kinetic proton temperature anisotropy
instabilities.\citep{Hellinger:2006,Bale:2009,Maruca:2011} The
physical cause for this spread of \tempAni values remains to be
definitively determined. 
For double adiabatic evolution of the temperatures in a magnetized
plasma,\cite{Chew:1956} the spherical expansion of the solar wind in
the inner heliosphere with a typical Parker spiral magnetic field
leads to \tempAni $<
1$.\cite{Hellinger:2003,Matteini:2005,Hellinger:2006,Matteini:2012} On
the other hand, mechanisms proposed to yield \tempAni $>1$ include
proton cyclotron heating,\citep{Hollweg:2002} shocks,\cite{Gary:1992}
compression between solar wind streams,\cite{Schwartz:1983},
compressional slow wave modes,\cite{TenBarge:2015} or double
adiabatic expansion with a transverse magnetic
field.\cite{Matteini:2012}
A discussion of the evolution of \tempAni with heliocentric radius can be found in
Matteini \emph{et. al.} 2007\cite{Matteini:2007}, and a review of the
many potential mechanisms governing \tempAni and their associated time
scales in the solar wind is presented in TenBarge \emph{et. al.}
2015\cite{TenBarge:2015}.  Here we will not further address the causes
of the proton temperature anisotropy, but merely accept it as an
observational fact and explore the consequences for the turbulence.

For a plasma with a bi-Maxwellian proton distribution and plasma
parameters relevant to the solar wind, there exist at least
four potential electromagnetic proton temperature anisotropy instabilities:
\textbf{(1)} the parallel (or whistler) firehose
instability;\citep{Kennel:1966,Gary:1976} \textbf{(2)} the \Alfven (or
oblique) firehose instability;\citep{Hellinger:2000} \textbf{(3)} the
mirror instability;\citep{Vedenov:1958,Tajiri:1967,Southwood:1993} and
\textbf{(4)} the proton cyclotron instability.\citep{Gary:1976} The
first two of these instabilities occur for \tempAni $<1$ and \Bpar
$>1$, while the latter two instabilities arise for \tempAni $>1$ and
all $\beta_{\parallel p}$, where \Bpar is the ratio of parallel proton
thermal pressure to magnetic pressure.  When the plasma exceeds a
threshold value of the proton temperature anisotropy, these
instabilities can tap the free energy associated with the anisotropic
proton temperature, driving electromagnetic fluctuations and
ultimately reducing the temperature anisotropy, thereby moving the
plasma back toward a state of marginal
stability.\citep{Boris:1977,Matteini:2006,Hellinger:2008}

It is important to emphasize that each of these instabilities is
necessarily associated with a normal wave mode, or eigenmode, of the
kinetic plasma.  Each solution of the linear kinetic dispersion
relation yields a complex eigenfrequency $\omega+i\gamma$, where the
sign of the imaginary component indicates whether the mode is growing
or damped. For a uniform plasma with an isotropic Maxwellian
distribution, all eigenmodes are damped. But if the proton temperature
is sufficiently anisotropic, unstable modes may grow, utilizing the
free energy in the velocity distribution. This leads to the injection
of energy into electromagnetic fluctuations in the solar wind plasma
at scales where the instability growth rate is positive. Here we aim
to make clear the connection between each instability and its
associated wave mode, to elucidate the properties of these
instability-driven modes, and to discuss the resulting contribution to
the fluctuations measured in the turbulent solar wind.

\subsection{Numerical Solution}
\label{sec:method}

To explore the properties of a plasma with a bi-Maxwellian proton
velocity distribution, we employ the numerical Vlasov-Maxwell linear
dispersion relation solver PLUME (Plasma in a Linear Uniform Magnetized Environment).
PLUME extends the solver described in
Quataert 1998\cite{Quataert:1998} by allowing for a bi-Maxwellian
equilibrium temperature distribution for both electrons and an
arbitrary number of ion species.  We have benchmarked our results
against the widely used linear dispersion relation solvers
WHAMP\cite{Roennmark:1982} and NHDS\cite{Verscharen:2013a} and found
agreement.

The general linear dispersion relation for a fully ionized,
proton-electron plasma with bi-Maxwellian particle distributions can
be expressed as
\begin{equation}
\omega_{VM}= \\
\omega(k_\perp \rho_p, k_\parallel d_p,
\beta_{\parallel p}, \frac{T_{\perp p}}{T_{\parallel p}}, 
\frac{T_{\perp e}}{T_{\parallel e}}, \frac{T_{\parallel p}}{T_{\parallel e}},
\frac{v_{t \parallel p}}{c}).
\end{equation}
The equilibrium magnetic field is $\V{B}_0= B_0 \hat{\mathbf{z}}$, and
we solve for the eigenfrequencies and eigenfunctions of a plane wave
with wavevector $\V{k}= k_\perp \hat{\mathbf{x}} + k_\parallel
\hat{\mathbf{z}}$.  The proton gyroradius and inertial lengths are
defined as $\rho_p= v_{t \perp p}/\Omega_p$ and $d_p= c/\omega_{pp}$,
the perpendicular and parallel temperatures (expressed in units of
energy) for species $s$ are $T_{\perp s}$ and $T_{\parallel s}$, the
proton thermal velocities parallel and perpendicular to $\V{B}_0$ are
$v_{t (\parallel, \perp) p}=\sqrt{2 T_{(\parallel, \perp) p} / m_p} $,
and the speed of light is $c$. The parallel proton plasma beta, or
ratio of parallel thermal to magnetic pressure, is $\beta_{\parallel
  p} = 8 \pi n T_{\parallel p}/ B_0^2$, and the proton gyrofrequency
and proton plasma frequency are given by $\Omega_p=q B_0/ (m_p c)$ and
$\omega_{pp}= \sqrt{4\pi n_p q^2/m_p}$.

In order to focus on the four proton temperature anisotropy
instabilities, we consider a non-relativistic plasma, $v_{t \parallel
  p}/c \ll 1$, in which the electrons have a Maxwellian equilibrium
distribution, $T_{\perp e}/T_{\parallel e} =1$, and the protons and
electrons have equal parallel temperatures, $T_{\parallel
  p}/T_{\parallel e}=1$.  The proton distribution is bi-Maxwellian,
with distinct perpendicular and parallel temperatures, $T_{\perp p}$
and $T_{\parallel p}$.  Under these assumptions, linear solutions of
the Vlasov-Maxwell dispersion relation are dependent on only four
parameters,
\begin{equation}
\omega_{VM}=\omega (k_\perp \rho_p, k_\parallel d_p, 
\beta_{\parallel p}, \frac{T_{\perp p}}{T_{\parallel p}}). 
\end{equation}

We avoid more complex distributions---such as a Kappa
distribution,\citep{Vasyliunas:1968,Gosling:1981,Christon:1988,Williams:1988}
distributions composed of several overlapping
Maxwellians,\cite{Marsch:1982,Marsch:2012} or multiple ion
species---because the non-Maxwellian nature of the bi-Maxwellian
solution depends only on the single parameter $T_{\perp
  p}/T_{\parallel p}$, providing physical insight into the fundamental
physics of these instabilities without the complications introduced
from more complicated non-Maxwellian distributions. Future work
concerning the alteration to the temperature anisotropy instabilities
from other non-Maxwellian
distributions,\citep{Xue:1993,Summers:1994,Xue:1996} additional ion
species,\cite{Podesta:2011b} and relative
drifts\cite{Matteini:2012,Verscharen:2013a,Verscharen:2013b} will
prove useful in extending the results of this work to achieve a more
complete characterization of the behavior of turbulent fluctuations in
the solar wind.

One of our main aims is to establish a clear connection between the
four proton temperature anisotropy instabilities and the low-frequency
normal modes of the Vlasov-Maxwell system, specifically the kinetic
counterparts of the MHD Alfv\'en, fast, slow, and entropy
modes.\cite{Klein:2012} Each of these four normal modes is associated
with a distinct dispersion surface.\cite{Andre:1985,Yoon:2008} A
dispersion surface is a map formed by the solution to the linear
dispersion relation (for example, the real component of the complex
eigenfrequency) over wavevector space, in this case the $(k_\perp,
k_\parallel)$ plane. Different regions in wavevector space on a single
dispersion surface correspond to different commonly known wave modes
that may have distinct properties.

To avoid confusion, we specify here each of these four dispersion
surfaces (associated with the Alfv\'en, fast, slow, and entropy modes)
and identify the commonly known wave modes represented by different
regions on each surface.  The \Alfven dispersion surface includes the
MHD \Alfven wave at $k_\perp \rho_p \ll 1$ and $k_\parallel d_p \ll
1$, the kinetic \Alfven wave at $k_\perp \rho_p \gtrsim 1$ and
$k_\parallel d_p \ll 1$, and the proton cyclotron wave at $k_\parallel
d_p \gtrsim 1$.  The fast dispersion surface includes the fast
magnetosonic wave at $k_\perp \rho_p \ll 1$ and $k_\parallel d_p \ll
1$, the ion Bernstein wave at $k_\perp \rho_p \gtrsim 1$ and
$k_\parallel d_p \ll 1$, and the whistler wave at $k_\parallel d_p
\gtrsim 1$. The associations between these commonly known wave modes
and the \Alfven and fast dispersion surfaces are illustrated in Figure
1 of Howes \emph{et al.}  2014.\cite{Howes:2014a} The slow and entropy
dispersion surfaces include the slow magnetosonic wave and
non-propagating entropy mode at $k_\perp \rho_p \ll 1$ and
$k_\parallel d_p \ll 1$; neither of these two dispersion
surfaces has commonly used names at the small, kinetic scales.  In the
remainder of this paper, we choose to identify the appropriate
dispersion surface by the associated large-scale, low-frequency mode:
the \Alfven wave, fast wave, slow wave, or entropy mode.
A detailed discussion regarding connecting MHD and Vlasov-Maxwell 
normal modes can be found in 
Krauss-Varban \emph{et al.} 1994\cite{Krauss-Varban:1994}
and Klein \emph{et al.} 2012.\cite{Klein:2012}

We employ the following rigorous procedure to connect each of the four
proton temperature anisotropy instabilities to one of the four
low-frequency, normal modes of the Vlasov-Maxwell system. Consider the
case that our numerical solver finds an unstable mode (a solution for
the complex eigenfrequency that has a positive imaginary component)
for a given set of the four parameters, $\mathcal{P}^*=[(k_\perp
  \rho_p)^*, (k_\parallel d_p)^*, \beta^*_{\parallel p}, (T_{\perp
    p}/T_{\parallel p})^*]$. First, we begin with the solutions for
each of the four normal wave modes for the initial parameters
$\mathcal{P}^0=(k_\perp \rho_p, k_\parallel d_p, \beta_{\parallel p},
T_{\perp p}/T_{\parallel p})= (10^{-3},10^{-3},1,1)$.  Next, we
perform a nearly continuous variation of each of these four parameters
until we reach the desired point in parameter space
$\mathcal{P}^*$. The nearly continuous variation for a given parameter
involves many small increments of that parameter; after each small
increment, the roots of the dispersion relation are recalculated,
using solutions from the previous parameters as initial guesses for
the new solutions.  Finally, the unstable mode is identified as one of
the four normal-mode solutions at $\mathcal{P}^*$.  This procedure
allows for a smooth connection between the well-established, isotropic
large-scale normal modes and the smaller scale, anisotropic modes and
instabilities.

Identifying these modes by such a continuous variation can be made
difficult by the presence of exceptional points on the solution
surface of the Vlasov-Maxwell dispersion
relation,\citep{Kato:1966,Heiss:2004a,Heiss:2004b,Kammerer:2008} a
technical point discussed in Appendix A.  For example, at
$\beta_{\parallel p}>1,$ the whistler wave is actually the extension
of the slow mode,\cite{Krauss-Varban:1994,Klein:2012} rather than the
fast mode, due to a large-scale mode exchange described in Appendix A.

\subsection{The Properties of the Proton Temperature Anisotropy Instabilities}
\label{sec:instable}

As discussed in Section~\ref{sec:method}, the linear dispersion
relation for a non-relativistic, fully ionized, proton-electron plasma
with a bi-Maxwellian proton distribution depends only on the four
parameters $(k_\perp \rho_p, k_\parallel d_p, \beta_{\parallel p},
T_{\perp p}/T_{\parallel p})$. Therefore, the two plasma parameters,
$\beta_{\parallel p}$ and $T_{\perp p}/T_{\parallel p}$, control
whether the plasma is unstable. In the event of instability, some
regions of wavevector space $(k_\perp \rho_p, k_\parallel d_p)$ will
have positive growth rates, where observations of the electromagnetic
fluctuations driven by the instability are generally expected to be
dominated by modes with wavevectors having the maximum growth rate.
The four proton temperature anisotropy instabilities arise in
different regions of $(\beta_{\parallel p}, T_{\perp p}/T_{\parallel
  p})$ plasma parameter space and lead to unstable wave growth in
distinct regions of $(k_\perp \rho_p, k_\parallel d_p)$ wavevector
space.  Determination of these regions provides useful information
concerning how these instabilities alter the existing large-scale
turbulent cascade and inject energy into the turbulence in the solar
wind, the key topics of this work.  In Section~\ref{sec:fluid}, we
classify the underlying fundamental mechanisms which control each of the
four proton temperature anisotropy instabilities, and use the
properties illustrated in Figures~\ref{fig:instable_map},
\ref{fig:parameter}, \ref{fig:cntr_omgam}, and \ref{fig:aleph_scan} to
examine the behavior of the four modes generated by these
instabilities as a function of the four parameters $(k_\perp \rho_p,
k_\parallel d_p, \beta_{\parallel p}, T_{\perp p}/T_{\parallel p})$.

\begin{figure}[t]
\hspace*{0.cm}
\includegraphics[width=7.0cm,viewport=25 15 150 150, clip=true]
{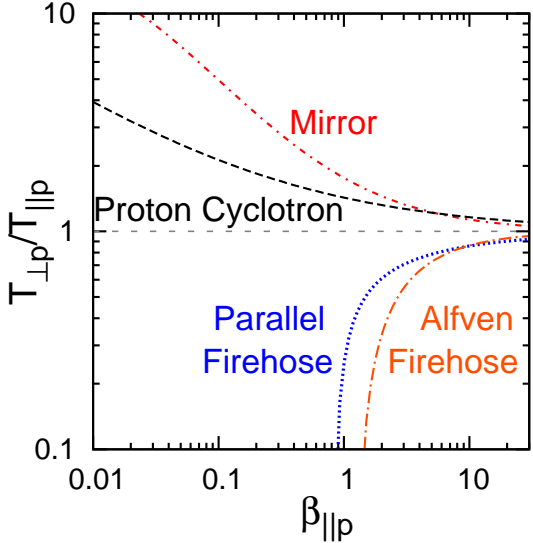}
\caption[Marginal stability conditions for the four ion temperature anisotropy
instabilities.]
{Marginal linear 
stability thresholds for the four proton temperature 
anisotropy instabilities using Equation~\ref{eq:unstableFit}
and the parameters in Table~\ref{tb:unstableFit}.
}
\label{fig:unstableFit}
\end{figure}

\begin{figure*}[t]
\begin{center}
\includegraphics[width=12.5cm,viewport=15 0 265 300, clip=true]
{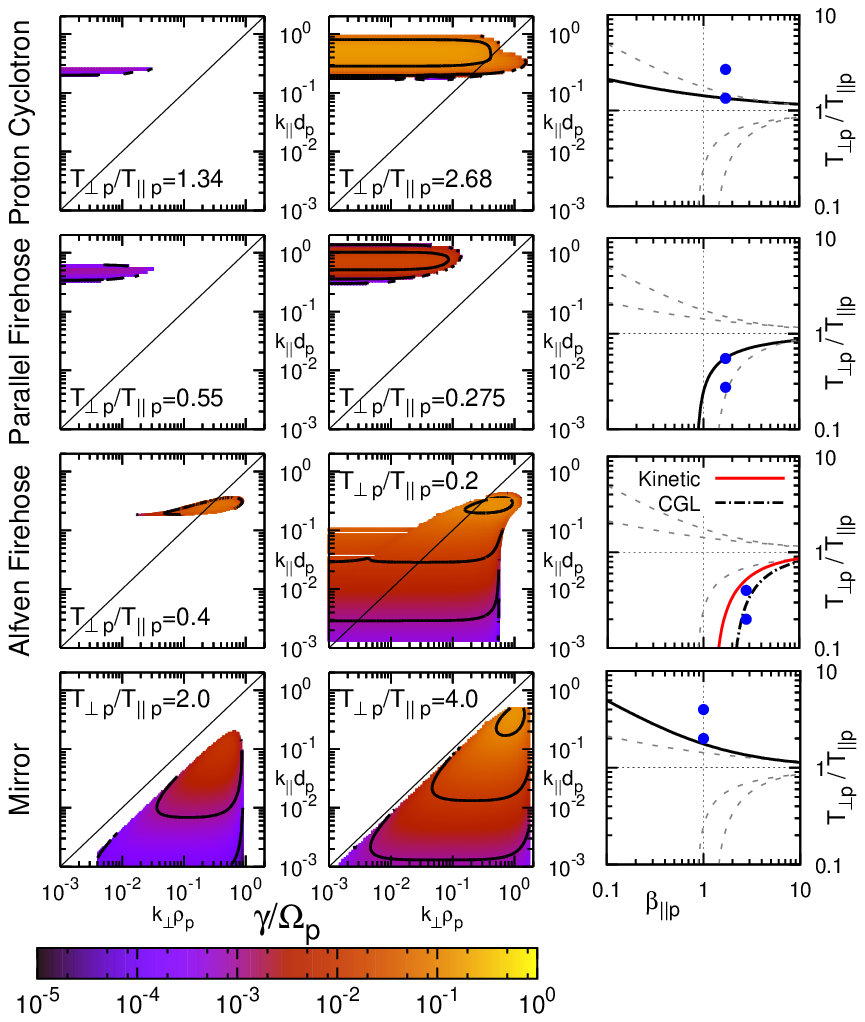}\\
\caption{ Maps in wavevector space of the growth rate
  $\gamma/\Omega_p$ for the unstable modes of the proton temperature
  anisotropy instabilities for fixed values of \Bpar and $T_{\perp
    p}/T_{\parallel p}$.  Both a marginally unstable (left column) and
  strongly unstable (center) case is presented for the proton
  cyclotron (first row), parallel firehose (second), \Alfven firehose
  (third), and mirror (fourth) instabilities. 
  A thin black line, indicating $k_\perp \rho_p = k_\parallel d_p$, is included
  as an aid in distinguishing between parallel and oblique unstable modes.
  In the right column are
  presented the points in $(\beta_{\parallel p}$,  $T_{\perp p}/T_{\parallel p})$ 
  parameter space from which these plots are
  generated as well as the associated marginal instability lines as
  calculated by Equation~\ref{eq:unstableFit}. }
\label{fig:instable_map}
\end{center}
\end{figure*}

\begin{figure*}[t]
\begin{center}
\hspace*{-1cm}
\includegraphics[width=14.5cm,viewport=0 0 335 310, clip=true]
{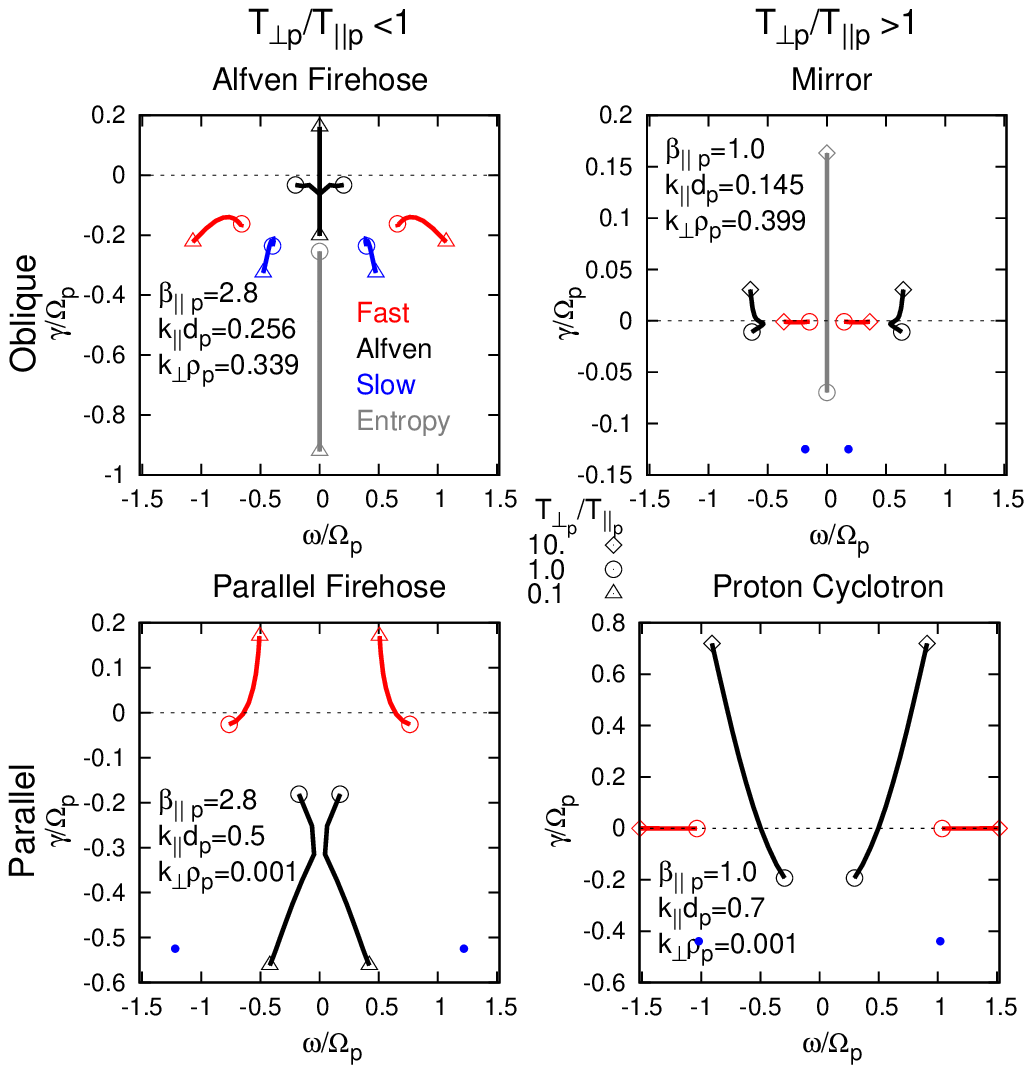}
\caption[Solutions for the Vlasov-Maxwell dispersion
relation given an anisotropic ion temperature distribution
parametrized by varying $T_{\perp p}/T_{\parallel p}$.]
{
Vlasov-Maxwell dispersion relation
solutions parametrized by a varying $T_{\perp p}/T_{\parallel p}$
for fast (red), \Alfven (black), slow  (blue), and entropy
(gray) modes in complex frequency space $(\omega/\Omega_p, \gamma/\Omega_p)$.
Solutions in each of the panels represent choices of plasma parameters
which will lead to the four proton temperature anisotropy instabilities:
the \Alfven firehose (top left), the mirror (top right), the parallel 
firehose (bottom left), and the proton cyclotron (bottom right).
\tempAni $<1$ $(>1)$ modes are shown in the left (right) column
and the oblique (parallel) modes are shown in the top (bottom)
row.
}
\label{fig:parameter}
\end{center}
\end{figure*}

\begin{figure*}[t]
\begin{center}
\includegraphics[width=13.20cm,viewport=5 0 500 145, clip=true]
{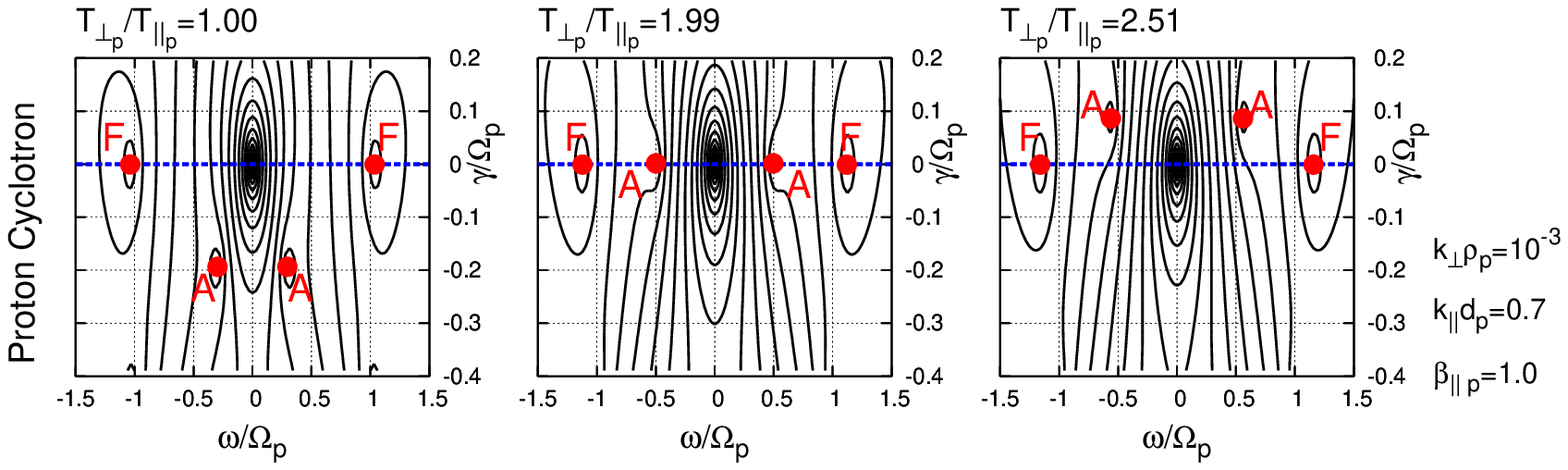}\\
\includegraphics[width=13.20cm,viewport=5 0 500 145, clip=true]
{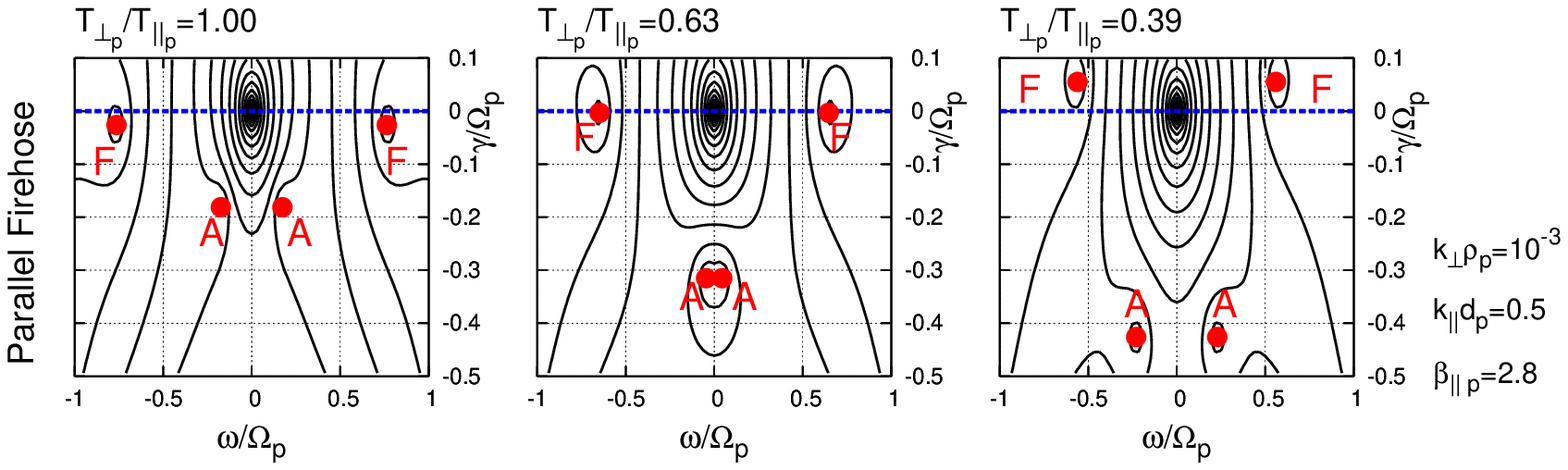}\\
\includegraphics[width=13.20cm,viewport=5 0 500 145, clip=true]
{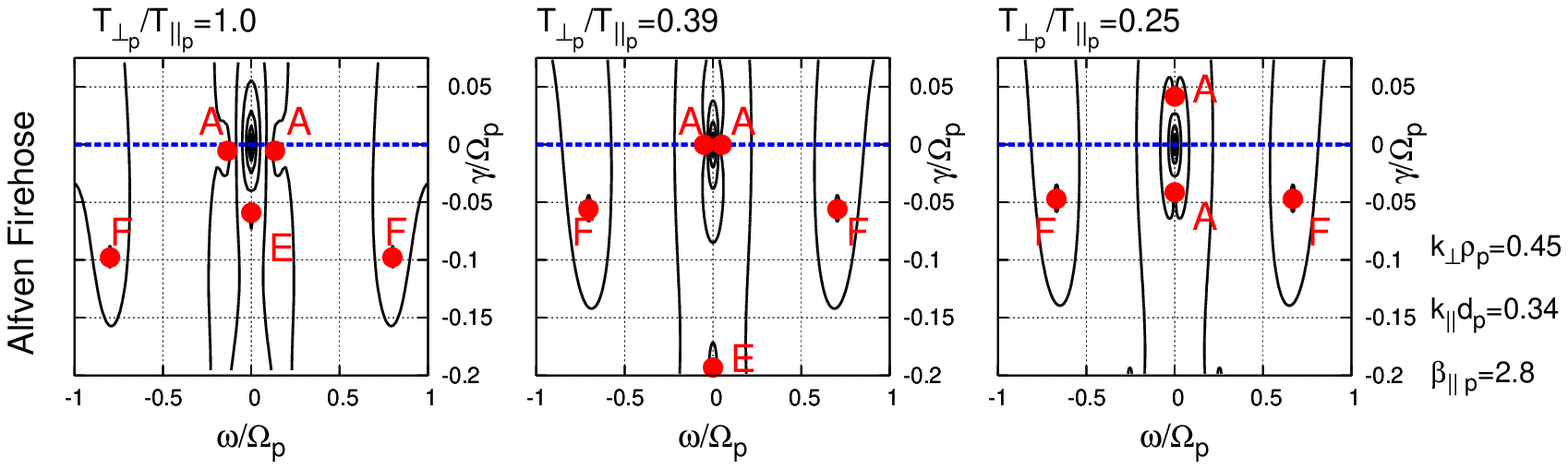}\\
\includegraphics[width=13.20cm,viewport=5 0 500 145, clip=true]
{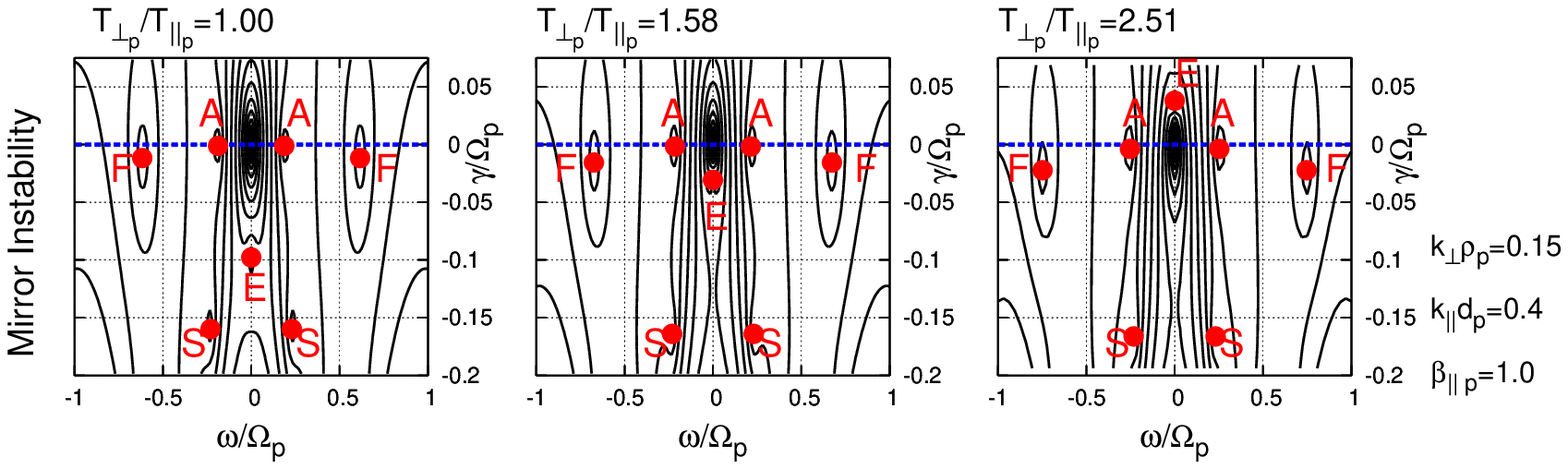}
\caption[]
{
Plots of the linear Vlasov-Maxwell dispersion 
relation $|\mathcal{D}(\omega/\Omega_p,\gamma/\Omega_p)|$
illustrating the transition 
of modes associated with the
proton temperature anisotropy instabilities from
stable (left column) to marginally stable (center)
to unstable (right). 
Contours of constant value are shown as black lines, with
eigenfrequency solutions for the Alfv\'en, fast, entropy,
and slow modes corresponding to 
$|\mathcal{D}(\omega/\Omega_p,\gamma/\Omega_p)|=0$ shown as red points.
The plasma parameters used for each contour map have been chosen to
generate the proton cyclotron (first row), parallel firehose (second),
\Alfven firehose (third), and mirror (fourth) instabilities.
The monotonic change of $T_{\perp p}/T_{\parallel p}$ is the only
variation of parameters made in each row. Discussions of the behavior of 
the instabilities are found in subsections~\ref{sec:fluid} A-D.
}
\label{fig:cntr_omgam}
\end{center}
\end{figure*}

\begin{figure*}[t]
\begin{center}
\includegraphics[width=13.00cm,viewport=10 15 480 180, clip=true]
{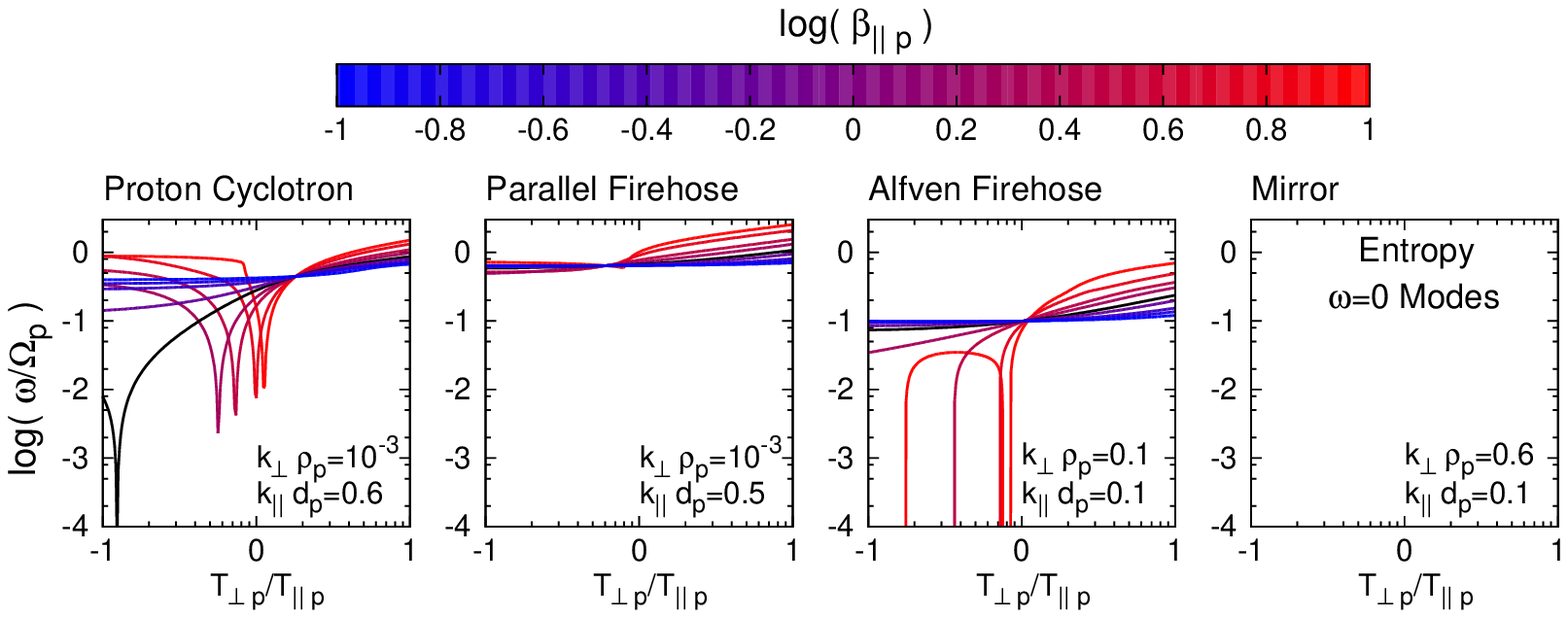}\\
\includegraphics[width=13.00cm,viewport=10 15 480 120, clip=true]
{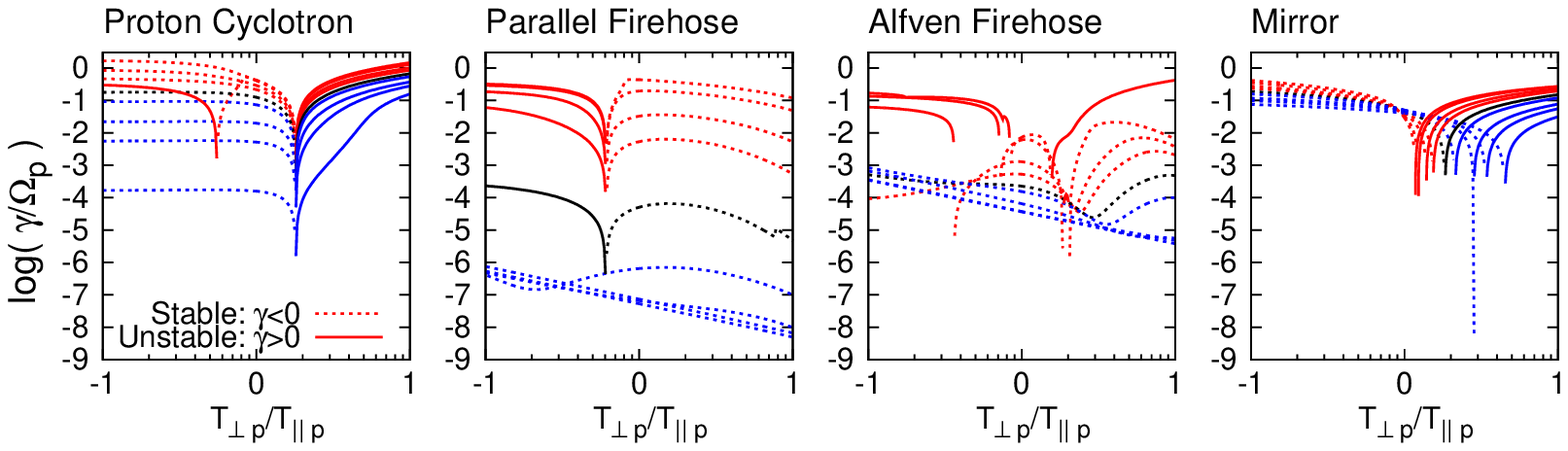}\\
\includegraphics[width=13.00cm,viewport=10 15 480 120, clip=true]
{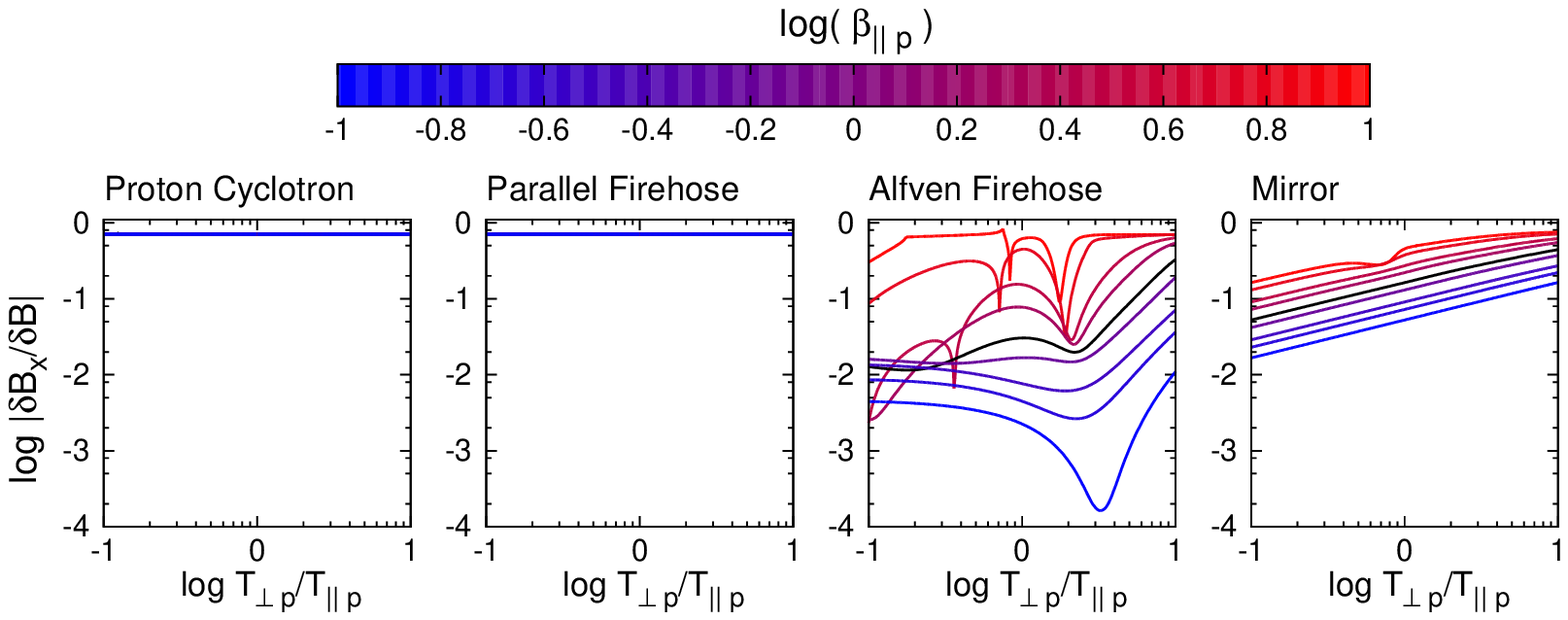}\\
\includegraphics[width=13.00cm,viewport=10 15 480 120, clip=true]
{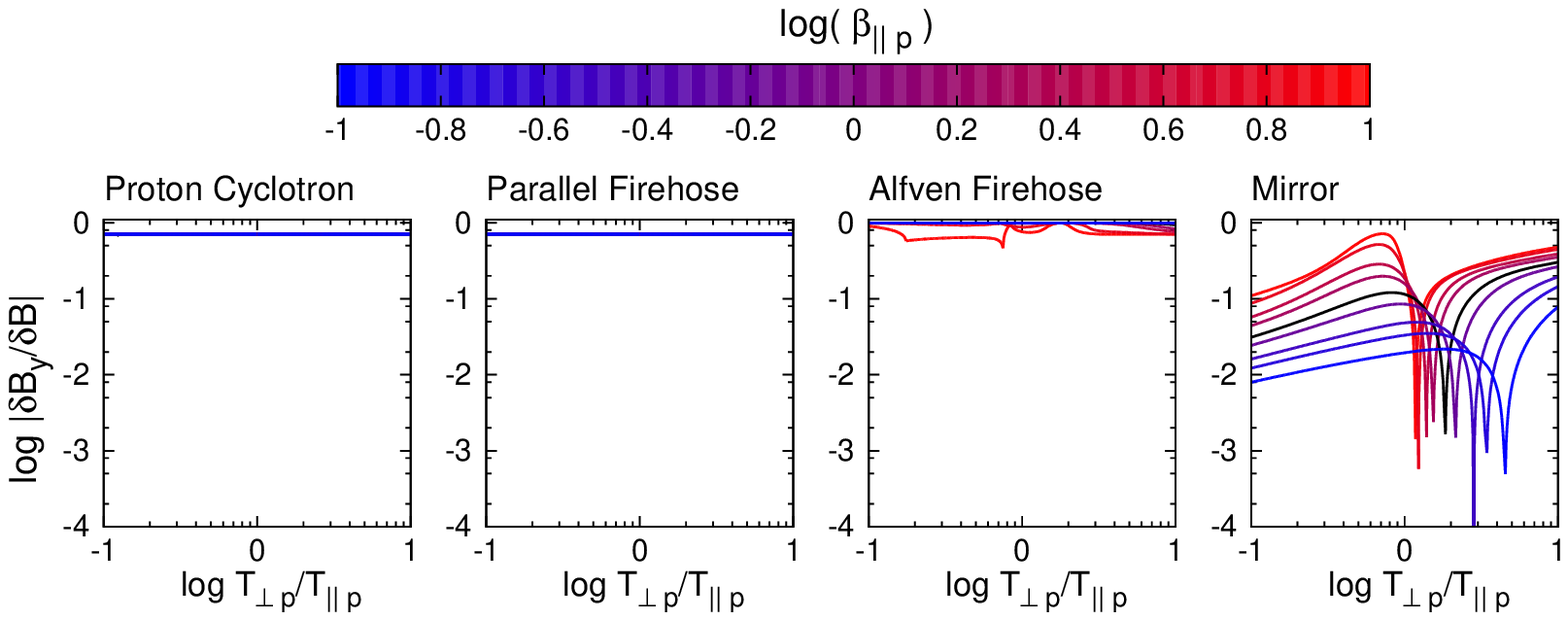}\\
\includegraphics[width=13.00cm,viewport=10 15 480 120, clip=true]
{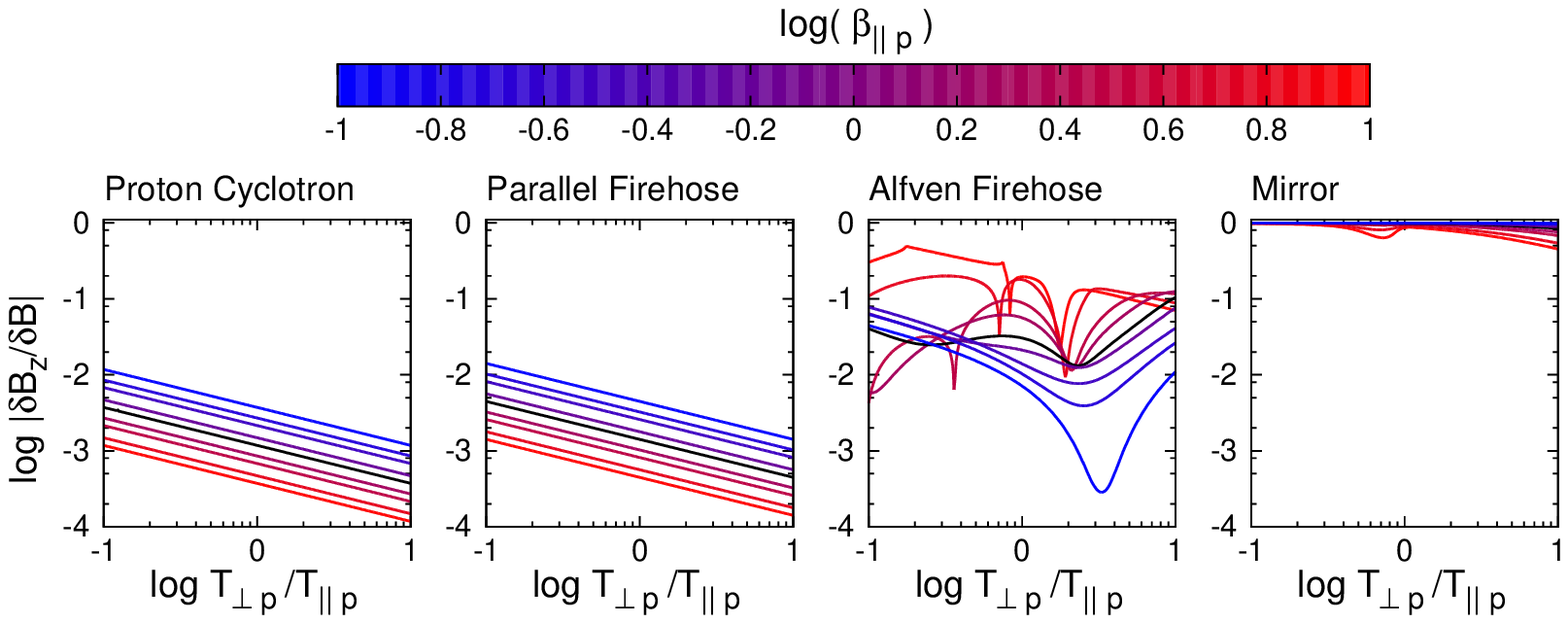}\\
\includegraphics[width=13.00cm,viewport=10 0 480 120, clip=true]
{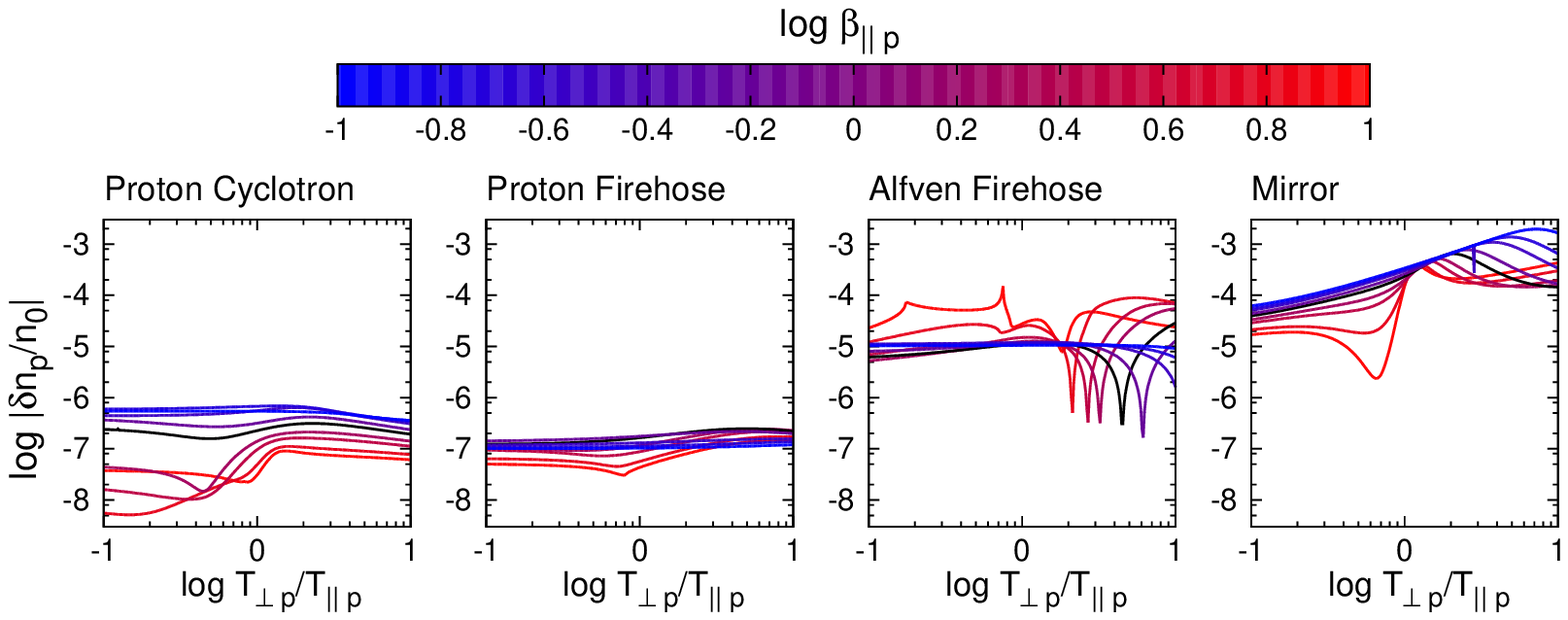}\\
\caption[]
{
Plots of $\omega/\Omega_p, \ \gamma/\Omega_p$, $\delta\V{B}/\delta B$,
and $\delta n/n_0$
(in descending rows)
for the four normal modes associated with proton temperature
anisotropy instabilities (columns) as a function of $T_{\perp p}/T_{\parallel p}$.
Values are plotted for two decades of $\beta_{\parallel p}$, ranging 
logarithmically from $0.1$ (blue) to $10.0$ (red), with $1.0$ given in black.
Each of the modes have fixed values of $k_\perp \rho_p$ and $k_\parallel d_p$.
In the second row, stable (unstable) modes with $\gamma <0$ ($>0$) 
are plotted with dashed (solid) lines.
}
\label{fig:aleph_scan}
\end{center}
\end{figure*}

Hellinger \emph{et al.} 2006\cite{Hellinger:2006} has compiled values
for the marginal stability boundaries for these instabilities in the
$(\beta_{\parallel p}, T_{\perp p}/T_{\parallel p})$ plane.  The
marginal stability boundary is set by calculating $\omega(\V{k})$ for
a fixed \Bpar over all possible wavevectors $\V{k}$.  The proton
temperature anisotropy \tempAni is then varied until the most unstable
wavevector has a growth rate of $\gamma/\Omega_p=10^{-3}$, thus
establishing the instability criterion.  These criteria are generally
well fit by an expression of the form
\begin{equation}
T_{\perp p}/T_{\parallel p}=1+\frac{a}{(\beta_{\parallel p}-\beta_0)^b}
\label{eq:unstableFit}
\end{equation} 
where $a, \ b,$ and $\beta_0$ 
are unique values for each of the four instabilities.
Values from Hellinger \emph{et al.} 2006 \cite{Hellinger:2006}
are found in Table~\ref{tb:unstableFit} and 
used to plot the four thresholds in 
Figure~\ref{fig:unstableFit}. 
Observations of \Bpar and \tempAni are 
approximately constrained by the mirror and \Alfven firehose boundaries
in the solar wind.\citep{Hellinger:2006,Bale:2009,Maruca:2011}

\begin{table}[t]
\caption{Instability Threshold Parameters for $\gamma/\Omega_i|_{max}=10^{-3}$ 
from Hellinger \emph{et al.} 2006\cite{Hellinger:2006}.}
\begin{center}
\begin{tabular}{|c|c|c|c|}
\hline
& $a$ & $b$ & $\beta_0$\\
\hline
Proton Cyclotron & 0.43 & 0.42 & -0.0004 \\
Parallel Firehose & -0.47 & 0.53 & 0.59 \\
\Alfven Firehose & -1.4 & 1.0 & -0.11 \\
Mirror & 0.77 & 0.76 & -0.016 \\
\hline
\end{tabular}
\end{center}
\label{tb:unstableFit}
\end{table}

For a single point in the $(\beta_{\parallel p}, T_{\perp
  p}/T_{\parallel p})$ plasma parameter space at which one of the
proton temperature anisotropy instabilities arises, the unstable modes
will only occupy a particular region of $(k_\perp \rho_p, k_\parallel
d_p)$ wavevector space. Generally, the parallel firehose and proton
cyclotron instabilities have their largest growth rates for nearly
parallel wavevectors with $k_\parallel d_p \sim 1$ and $k_\perp \rho_p
\ll 1$, while the mirror and \Alfven firehose are most unstable at
oblique angles with $k_\parallel d_p \sim k_\perp \rho_p \sim 1$.  To
illustrate the unstable regions of $(k_\perp \rho_p, k_\parallel d_p)$
wavevector space for each of the four instabilities, we present in
Figure~\ref{fig:instable_map} maps of the positive growth rates
$\gamma/\Omega_p$ for the modes associated with the four instabilities
in $(k_\perp \rho_p, k_\parallel d_p)$ wavevector space at fixed
values of \Bpar and $T_{\perp p}/T_{\parallel p}$.  A discussion of
the classification of microscale instabilities, with $k_\perp \rho_p
\simeq 1$ and/or $k_\parallel d_p \simeq 1$, and macroscale
instabilities, with $k_\perp \rho_p \ll 1$ and $k_\parallel d_p \ll
1$, is presented in Section~\ref{sec:fluid}.  Note that the relation
between the proton inertial length and gyroradius depends on both the
parallel plasma beta and the proton temperature anisotropy,
$\rho_p=d_p \sqrt{\beta_{\parallel p} T_{\perp p}/T_{\parallel p}}$.

To visualize the connection between each proton temperature anisotropy
instability and its associated linear eigenmode, we present in
Figure~\ref{fig:parameter} solutions for the fast (red), \Alfven
(black), slow (blue), and entropy (gray) modes in complex frequency
space $(\omega/\Omega_p,\gamma/\Omega_p)$ parametrized as a function
of $T_{\perp p}/T_{\parallel p}$.  For each panel, we chose
appropriate values of $\beta_{\parallel p}$, $k_\perp \rho_p$ and
$k_\parallel d_p$ to allow for the development of one of the proton temperature
anisotropy instabilities and vary \tempAni from unity, indicated by an
open circle, to $T_{\perp p}/T_{\parallel p}=10$ ($T_{\perp
  p}/T_{\parallel p}=0.1$) indicated with an open diamond (triangle)
in the right (left) panels.  The oblique instabilities, with $k_\perp
\rho_p \sim k_\parallel d_p \sim 1$, are plotted in the top row and
the parallel instabilities, with $k_\perp \rho_p \ll k_\parallel d_p
\sim 1$, are plotted in the bottom row.

Figure~\ref{fig:parameter} shows clearly that the proton cyclotron
instability (lower right) is connected to the proton cyclotron wave
(black), and is therefore associated with the \Alfven wave.  The
mirror instability (upper right) is connected to the non-propagating
entropy mode (grey).  The parallel firehose instability (lower left)
is connected to the whistler wave (red), and is therefore associated
with the fast wave. Finally, the \Alfven firehose instability (upper
left) is connected to non-propagating \Alfvenic fluctuations (black).
It should be noted that the unstable \Alfven wave solution (shown in
black) for the mirror instability parameters (upper right) is due to
an extension of the proton cyclotron instability to oblique angles for
the very anisotropic temperature ratio \tempAni $\simeq 10$, and is not
related to the mirror instability.  The behavior of each of these
parametrized paths, a key result of this paper,
 will be discussed thoroughly in Section~\ref{sec:fluid}.

Figure~\ref{fig:cntr_omgam} plots the roots of the Alfv\'en, fast, slow,
and entropy modes in complex frequency space $(\omega/\Omega_p,\gamma/\Omega_p)$
for fixed values of $\beta_{\parallel p}$, $k_\parallel d_p$, and $k_\perp \rho_p$.
The values of these three parameters are distinct for each instability,
and have been chosen to maximize the growth of the associated unstable mode.
The instabilities are ordered by row as proton cyclotron (top),
parallel firehose (second), \Alfven firehose (third), and   
mirror (bottom).
The proton temperature anisotropy is varied from isotropy (left column)
to a value near marginal stability (center), 
to a value of unstable growth (right).  
Also plotted are contours for the amplitude of the linear dispersion relation 
$|\mathcal{D}(\omega,\gamma)|$ (c.f. Equation 10-73 in Stix 1992 \citep{Stix:1992}), 
where solutions to the linear dispersion relation are found at points 
where $|\mathcal{D}|=0$.
These contours, while not physically meaningful, do provide a
guide for the eye to where the solutions of the 
linear dispersion relation occur in complex frequency space.

In Figure~\ref{fig:aleph_scan}, we plot the complex eigenfrequencies
and magnetic and density eigenfunctions for the four modes associated
with the proton temperature anisotropy instabilities as a function of \tempAni and
$\beta_{\parallel p}$ at fixed points in $k_\parallel d_p$ and
$k_\perp \rho_p$, showing the eigenfunction characteristics for both
stable and unstable values of the plasma parameters $(\beta_{\parallel
  p}, T_{\perp p}/T_{\parallel p})$.  As with
Figure~\ref{fig:cntr_omgam}, a single wavevector $(k_\perp \rho_p,
k_\parallel d_p)$ is chosen for which unstable modes exist for some
values of $\beta_{\parallel p}$ and $T_{\perp p}/T_{\parallel p}$.
The instabilities are plotted by column as proton cyclotron (first),
parallel firehose (second), \Alfven firehose (third), and mirror
(fourth).  The parameters \tempAni and $\beta_{\parallel p}$ are both
varied from $0.1$ to $10$, with the \tempAni variation shown on the
horizontal axis and the \Bpar variation shown in color, with blue
indicating $\beta_{\parallel p}=0.1$, black indicating \Bpar $=1$, and
red indicating $\beta_{\parallel p}=10$.  By row from the top, we plot
$\omega/\Omega_p$ (first), $|\gamma/\Omega_p|$ (second), $|\delta
B_x/\delta B|$ (third), $|\delta B_y/\delta B|$ (fourth), $|\delta
B_z/\delta B|$ (fifth), and $|\delta n_p/n_0|$ (sixth).  In the
second row, the unstable (stable) modes are indicated by solid
(dashed) lines for $\gamma/\Omega_p >0 \ (<0)$.  


\section{Taxonomy of Proton Temperature Anisotropy Instabilities}
\label{sec:fluid}

The classification of plasma instabilities is an important but often
subtle matter.  A variety of different schemes have been applied,
contrasting fluid vs.~kinetic instabilities, macroinstabilities
vs.~microinstabilities, or configuration-space vs.~velocity-space
instabilities. These different schemes classify the instabilities
according to different criteria, such as the spatial scales at which
the instability operates or the nature of the underlying mechanism
driving the instability.  For instance, Treumann \& Baumjohann
1997\cite{Treumann:1997} categorize instabilities as either
macroinstabilities or microinstabilities depending on the spatial
scales associated with the instability.  Macroinstabilities occur at
scales much larger than the particle kinetic scales ($\rho_s$ or
$d_s$), while microinstabilities arise near these kinetic scales.
Krall \& Trivelpiece 1973\cite{Krall:1973} alternatively define
configuration-space instabilities to be those which are associated
with the departure of macroscopic quantities from thermodynamic
equilibrium, while velocity-space instabilities are those which depend
on departures from an isotropic Maxwellian velocity distribution
function.  The configuration-space vs.~velocity-space distinction is
also often referred to as a fluid vs.~kinetic distinction.  A detailed
discussion about the relation between a plasma's macroscopic
quantities and associated microscopic instabilities can be found in
Schekochihin \emph{et al.}  2010\cite{Schekochihin:2010}.

It must be noted that the Treumann and Krall definitions are not
synonymous as they are based on different criteria; there exist
instabilities which occur at large spatial scales which are driven by
velocity-space effects.  These instabilities are macroscopic by the
Treumann definition and velocity-space instabilities by the Krall
definition.  The mirror instability is such an instability, which
arises across a broad spectrum of scales (see the bottom row of
Figure~\ref{fig:instable_map}), but whose mechanism relies on a
resonant response of a narrow region of the velocity distribution
function.\cite{Southwood:1993} Here we use the macroscopic/microscopic
terminology solely to classify the spatial scales at which the
instability arises and the configuration/velocity space terminology to
classify the nature of the instability mechanism.

The double adiabatic Chew-Goldberger-Low\citep{Chew:1956} (CGL)
firehose instability is a canonical example of a macroscopic,
configuration-space instability.  The instability arises for large
spatial scales with $k \rho_p \ll 1$, as shown in row three, center
column of Figure~\ref{fig:instable_map}, for $\beta_\parallel >1$ and
$T_\perp/T_\parallel<1$.  The mechanism driving the instability, that
the perpendicular pressure is not sufficient to counteract the
centrifugal force from a bent magnetic flux tube, relies only on the
macroscopic quantities of perpendicular and parallel pressure, and
therefore depends on $\beta_\parallel$ and $T_\perp/T_\parallel<1$
(the single-fluid analogs of \Bpar and $T_{\perp p}/T_{\parallel
  p}<1$).  Though the parallel firehose and the \Alfven firehose
instabilities occur in the same region of $(\beta_{\parallel p},
T_{\perp p}/T_{\parallel p})$ parameter space, they are both
fundamentally different from the CGL instability.  Both are
microscopic, velocity-space instabilities which depend on the
resonance condition $\omega(\V{k}) - k_\parallel v_\parallel \pm
\Omega_p = 0$ and not on the bulk, thermodynamic quantities of
perpendicular and parallel pressure.  The wavevectors for which the
microscopic firehose instabilities arise are shown in the second and
third rows of Figure~\ref{fig:instable_map}.  Both the parallel and
\Alfven firehose instabilities have marginal stability thresholds
closer to temperature isotropy than the CGL firehose, meaning that for
a plasma with \tempAni decreasing from unity, the velocity-space
instabilities will arise before the configuration-space instability.
 
\subsection{Proton Cyclotron Instability}
\label{sec:pci}

The proton cyclotron instability is a microscopic, velocity-space
instability driven by the resonant condition $\omega(\V{k}) -
k_\parallel v_\parallel - \Omega_p = 0$.  This instability couples to
the proton cyclotron wave,\citep{Kennel:1966, Davidson:1975} the
extension of the \Alfven wave at small parallel scales, $k_\parallel
d_p \gtrsim 1$. It occurs for \tempAni $>1$ and for all
$\beta_{\parallel p}$, with the marginal stability threshold
decreasing with increasing $\beta_{\parallel p}$.

The top row of Figure~\ref{fig:cntr_omgam} presents the fast and
\Alfvenic roots in complex frequency space for plasma parameters
relevant to the proton cyclotron instability: $\beta_{\parallel
  p}=1.0$, $k_\perp \rho_p=10^{-3}$, $k_\parallel d_p=0.7$, and
$T_{\perp p}/T_{\parallel p} =1.00$ (left column), $1.99$ (center),
and $2.51$ (right).  The slow and entropy modes are too heavily damped
to be shown in these plots.  For \tempAni $=1$, the proton cyclotron
waves are damped at a rate of $\gamma/\Omega_p\approx -0.2$.  Once the
perpendicular temperature is equal to twice the parallel, the mode
transitions from damped to growing, with increased anisotropy leading
to an increased growth rate of $\gamma/\Omega_p \simeq 0.1$ at
\tempAni $=2.5$. The damping of the fast waves is largely unaffected
by the presence of the anisotropy, with a slight increase in the real
frequency of the fast waves.

The microscopic nature of the proton cyclotron instability is shown in
the top row of Figure~\ref{fig:instable_map}, which plots the growth
rate $\gamma/\Omega_p>0$ as a function of wavevector for unstable
modes with \Bpar $=1.7$ and \tempAni $=1.344$ (left column) and
\tempAni $=2.688$ (center column).  These parameters, indicated as
blue dots (right column), are compared to the associated instability
threshold (black line) in the $(\beta_{\parallel p},T_{\perp
  p}/T_{\parallel p})$ plane.  The instability is confined to $0.1 <
k_\parallel d_p < 1$ for both values of $T_{\perp p}/T_{\parallel p}$.
While the unstable region for the marginally unstable case is
restricted to a nearly parallel region with $k_\perp \rho_p \ll 1$,
the unstable wavemodes for the highly unstable case include more
oblique modes with $k_\perp \rho_p \lesssim 1.0$.  It should be noted
that the position in the ($\beta_{\parallel p}$, \tempAni) plasma
parameter space for the latter case is not typically observed in the
solar wind (presumably because it is highly unstable).

The first column of Figure~\ref{fig:aleph_scan} illustrates the
$(\beta_{\parallel p},T_{\perp p}/T_{\parallel p})$ frequency and
eigenfunction dependence of the proton cyclotron wave for $k_\perp
\rho_p = 10^{-3}$ and $k_\parallel d_p = 0.6$.  The frequency of the
wave (top row) is fairly well constrained by $0.1 \Omega_p \lesssim
\omega \lesssim \Omega_p$, with some exceptions for portions of the
\Bpar $>1$, \tempAni $<1$ curves.  The damping rate $\gamma/\Omega_p$
(second row) of the stable modes is largely insensitive to \tempAni
but is strongly dependent on $\beta_{\parallel p}$, while the
transition from damping to growth is only dependent on $T_{\perp
  p}/T_{\parallel p}$. The wave is left-hand circularly polarized
with $|\delta B_x| \simeq |\delta B_y|$ and $|\delta B_z| \ll |\delta
B_y|$ (third through fifth rows).  This circular polarization is
utterly insensitive to changes in either \Bpar or $T_{\perp
  p}/T_{\parallel p}$ over the wide range plotted.  The mode is also
nearly incompressible, with no significant increases in
compressibility due to deviations from isotropy (sixth row).  Note
that the \Bpar $=10$ (red), \tempAni $<1$ unstable mode (second row)
is due to exceptional points of the type discussed in Appendix A, and
is a manifestation of the parallel firehose instability rather than the
proton cyclotron instability.

\subsection{Parallel Firehose Instability}
\label{sec:pfi}

The parallel firehose instability is a microscopic, velocity-space
instability driven by the resonant condition $\omega(\V{k}) -
k_\parallel v_\parallel + \Omega_p = 0$.  This instability couples to
the whistler wave,\citep{Quest:1996,Gary:1998} the extension of the
fast magnetosonic wave at small parallel scales, $k_\parallel d_p
\gtrsim 1$.  The parallel firehose instability occurs for \Bpar $>1$
and \tempAni $<1$, with the marginal stability threshold decreasing
for increasing $\beta_{\parallel p}$.

The second row of Figure~\ref{fig:cntr_omgam} presents the fast and
\Alfven roots in complex frequency space for plasma parameters
relevant to the parallel firehose instability: \Bpar $=2.8$, $k_\perp
\rho_p = 10^{-3}$, $k_\parallel d_i = 0.5$, and \tempAni $=1.00$ (left
column), $0.63$ (center), and $0.39$ (right).  As with the proton
cyclotron instability plots, the slow and entropy modes are too
heavily damped to be shown.  The fast mode passes through marginal
stability at $T_{\perp p}/T_{\parallel p}=0.63$ and becomes more
unstable for smaller $T_{\perp p}/T_{\parallel p}$.  The \Alfven root
represents the proton cyclotron wave, which becomes more strongly
damped with decreasing $T_{\perp p}/T_{\parallel p}$. In addition, the
real frequency of the fast mode decreases slightly with decreasing
$T_{\perp p}/T_{\parallel p}$.

The microscopic nature of the parallel firehose instability is shown
in the second row of Figure~\ref{fig:instable_map}, which plots the
growth rate $\gamma/\Omega_p>0$ as a function of wavevector for
unstable modes with \Bpar $=1.7$ and \tempAni $=0.55$ (left column)
and \tempAni $=0.275$ (center column).  These parameters, indicated as
blue dots (right column), are compared to the associated instability
threshold (black line) in the $(\beta_{\parallel p},T_{\perp
  p}/T_{\parallel p})$ plane.  As with the proton cyclotron
instability, unstable modes for the parallel firehose instability are
confined to $0.1 < k_\parallel d_p < 1$.  The perpendicular extent of
the instability does not significantly grow with smaller $T_{\perp
  p}/T_{\parallel p}$, with a maximum perpendicular scale around
$k_\perp \rho_p \simeq 0.1$, yielding unstable wavevectors within an
angle $\theta \simeq \tan^{-1}(0.1) \simeq 6^\circ$ from the
equilibrium magnetic field $\V{B}_0$.  The growth rate increases with
decreasing $T_{\perp p}/T_{\parallel p}$, but it never significantly
exceeds $\gamma/\Omega_p =0.05$.  
Although the extent of the unstable region in $(k_\perp
\rho_p,k_\parallel d_p)$ wavevector space for the parallel firehose
instability is similar to that of the other parallel instability, the
proton cyclotron instability (compare first and second rows), for the
more unstable cases (center column) the proton cyclotron instability
extends to more oblique angles of $\theta \lesssim 45^\circ$, rather
than the limit of $\theta \lesssim 6^\circ$ for the parallel firehose
instability.

The second column of Figure~\ref{fig:aleph_scan} illustrates the
$(\beta_{\parallel p},T_{\perp p}/T_{\parallel p})$ frequency and
eigenfunction dependence of the parallel whistler wave for $k_\perp
\rho_p = 10^{-3}$ and $k_\parallel d_p = 0.5$.  The frequency (top
row) is very tightly constrained to $\omega/\Omega_p \sim 1$
regardless of $\beta_{\parallel p}$ or $T_{\perp p}/T_{\parallel p}$.
There is a significant dependence of $\gamma/\Omega_p$ (second row) on
\tempAni for \Bpar $>1$.  The parallel whistler waves driven by the
parallel firehose instability are right-hand circularly polarized,
with $|\delta B_x| \simeq |\delta B_y|$ and $|\delta B_z| \ll |\delta
B_y|$ (third through fifth rows), and are nearly incompressible (sixth
row).  These magnetic and density eigenfunctions of the whistler waves
associated with the parallel firehose instability are insensitive to
changes in \Bpar or $T_{\perp p}/T_{\parallel p}$.

\subsection{\Alfven Firehose Instability}
\label{sec:afi}

The \Alfven firehose instability is a microscopic, velocity-space
instability driven by the resonant condition $\omega(\V{k}) -
k_\parallel v_{\parallel} \pm \Omega_p = 0$.\cite{Hellinger:2000}  This
instability couples to non-propagating, $\omega = 0$ oblique \Alfven
waves, and occurs for \Bpar $>1$ and \tempAni $<1$. The zero real
frequency region of the \Alfven dispersion surface grows with
increasing \Bpar and decreasing $T_{\perp p}/T_{\parallel p}$,
transforming the resonant condition to $k_\parallel = \pm \Omega_p/v_{\parallel}$.

The third row of Figure~\ref{fig:cntr_omgam} presents the Alfv\'en,
fast, and entropy roots in complex frequency space for parameters
relevant to the \Alfven firehose instability: \Bpar $=2.8$, $k_\perp
\rho_p=0.45$, $k_\parallel d_p=0.34$, and \tempAni $=1$ (left column),
$0.39$ (center), and $0.25$ (right).  The real frequency of the
propagating \Alfven waves decreases for decreasing \tempAni until the
$\pm \omega$ modes become degenerate at $\omega=0$ at the marginally
stable state.  A further decrease in \tempAni breaks the degeneracy of
these two \Alfven solutions, causing one of the modes to damp and the
other to grow unstable.  Both stable and unstable modes remain
non-propagating with real frequency $\omega=0$. As the nonlinear
saturation of the instability pushes the plasma from an unstable point
in $(\beta_{\parallel p}, T_{\perp p}/T_{\parallel p})$ parameter
space back towards a state of marginal stability, the
instability-driven, non-propagating modes transition back into
propagating oblique \Alfven waves.\cite{Hellinger:2008}

The wavevector regions of unstable growth for the \Alfven and CGL
firehose instabilities are shown in the third row of
Figure~\ref{fig:instable_map}.  Plasma parameters are set to
$\beta_{\parallel p}=2.8$ and $T_{\perp p}/T_{\parallel p}=0.4$ (left
column) and $T_{\perp p}/T_{\parallel p}=0.2$ (center).  These
parameters, indicated as blue dots (right column), are plotted against
the \Alfven (red solid) and CGL (black dash-dot) firehose instability
thresholds.  For the marginally unstable plasma (left), only the
microscopic \Alfven firehose instability arises, with the unstable
region tightly constrained to a narrow teardrop-shaped region near
$k_\perp \rho_p \sim k_\parallel d_p \sim 1$.  For the more unstable
case with \tempAni $=0.2$ (center), both the microscopic \Alfven and
macroscopic CGL instabilities occur.  The CGL firehose instability
causes a broad spectrum of unstable modes with $k_\parallel d_p < 0.1$
and $k_\perp \rho_p<0.6$, while the \Alfven firehose is still
restricted to the oblique wavevectors at kinetic scales with $k_\perp
\rho_p \sim k_\parallel d_p \sim 1$.  The CGL firehose unstable modes
have a smaller growth rate than the \Alfven firehose unstable modes,
but the unstable CGL modes are very widely distributed in wavevector
space, potentially allowing them to dynamically interact with the
anisotropic fluctuations of the large-scale turbulent cascade.  

The third column of Figure~\ref{fig:aleph_scan} illustrates the
$(\beta_{\parallel p},T_{\perp p}/T_{\parallel p})$ frequency and
eigenfunction dependence of the oblique kinetic scale \Alfven wave for
$k_\perp \rho_p = 0.1$ and $k_\parallel d_p = 0.1$.  For \Bpar $\leq
1,$ the real frequency (top row) $\omega/\Omega_p \simeq 0.1$, while
for increasing $\beta_{\parallel p}$, $\omega/\Omega_p$ increases for
\tempAni $>1$ and decreases for \tempAni $<1$, asymptoting to $\omega
= 0$ in the unstable region.  Note that for \Bpar $=10$ (red), the
\Alfven mode transitions between propagating ($\omega\ne 0$) and
non-propagating ($\omega = 0$) several times as \tempAni decreases
from $1$ to $0.1$. The \Bpar $=10$, \tempAni $>1$ unstable mode
(second row) is the proton cyclotron instability, which reaches
oblique angles for very high degrees of anisotropy.  The magnetic
eigenfunction (third through fifth rows) for the stable \Alfven wave
(with \Bpar$\le 1$) is dominated by $\delta B_y$, with a transition to
elliptical polarization for the \Alfven firehose unstable modes at
$\beta_{\parallel p}> 1$.  While more compressible than the two
parallel unstable modes, the oblique \Alfven wave is still mostly
incompressible (sixth row). The magnetic eigenfunctions associated
with the \Alfven firehose instability have significantly more
dependence on \Bpar and \tempAni than those associated with the
parallel instabilities.

\subsection{Mirror Instability}
\label{sec:mir}

The mirror instability is a macroscopic, velocity-space instability
which couples to oblique entropy
modes.\cite{Tajiri:1967,Southwood:1993} It occurs for \tempAni $>1$
and for all $\beta_{\parallel p}$, with the marginal stability
threshold decreasing for increasing $\beta_{\parallel p}$.

The mirror instability was originally proposed as a
configuration-space instability\cite{Rudakov:1961,Thompson:1964} which
was triggered by the anti-phase correlation between macroscopic
changes in the pressure and magnetic field strength. The instability
was shown to couple with a non-propagating mode, though it was thought
that the associated stable mode was propagating.  This mode was
identified as the MHD slow wave. Kinetic descriptions 
\cite{Vedenov:1958,Tajiri:1967} have demonstrated that the
physics of the configuration-space description was incorrect, due to
its inability to model the velocity-space effects driving the
instability.  Unfortunately, the conception of the mirror instability
as a configuration-space instability has persisted in the literature
for several decades.  The instability threshold calculated from
configuration-space theory is still correct, but the mode driven
unstable is the entropy mode, not the slow wave.\cite{Southwood:1993}

The mirror instability actually arises due to the difference between
the anti-phase response of the bulk plasma's thermal pressure 
to magnetic pressure perturbations and the in-phase response of 
particles with $v_\parallel \simeq 0$. 
The particles with $v_\parallel \simeq 0$, typically referred to as resonant particles,
move very little along the magnetic field lines, gaining or losing
energy with increasing or decreasing field strength. Particles in 
the bulk of the plasma with finite $v_\parallel$ stream along the magnetic
field lines, largely conserving particle energy by transferring it
between their parallel and perpendicular degrees of freedom.
A through review and discussion of the physics relevant to the mirror
instability can be found in Southwood \& Kivelson 1993.\cite{Southwood:1993}

The marginal stability threshold occurs at larger \tempAni for the
mirror instability than for the proton cyclotron instability (see
Figure~\ref{fig:unstableFit}), leading to the
contention\citep{Gary:1992} that the larger proton temperature
anisotropies needed to drive the mirror instability would be
isotropized by the proton cyclotron instability before the mirror
instability could be triggered.  However, the inclusion of minor
ions\citep{Price:1986} can alter the instability thresholds, reducing
the linear growth rate of the proton cyclotron instability while
leaving the mirror instability mostly unchanged. Additionally, solar
wind measurements of $(\beta_{\parallel p},T_{\perp p}/T_{\parallel
  p})$ usually appear to be constrained by the
marginal stability threshold of the mirror instability rather than
that of the proton cyclotron
instability.\citep{Hellinger:2006,Bale:2009} Which of these two
instabilities dominates in the free-streaming solar wind and what
mechanisms control their interaction remains an open scientific
question.

The bottom row of Figure~\ref{fig:cntr_omgam} presents the Alfv\'en,
fast, slow, and entropy roots in complex frequency space for
parameters relevant to the mirror instability: \Bpar $=1.0$, $k_\perp
\rho_p=0.15$, $k_\parallel d_p=0.4$, and \tempAni $=1.0$ (left
column), $1.58$ (center), and $2.51$ (right).  For these parameters,
the entropy mode is less damped than the slow modes at \tempAni
$=1.0$, and as $T_{\perp p}/T_{\parallel p}$ increases, the damping
rate of the entropy mode decreases to zero, reaching the marginally
stable state at $T_{\perp p}/T_{\parallel p}=1.58$.  Above this value,
the mirror instability is triggered, resulting in a growing,
non-propagating ($\omega=0$) mode. Like the \Alfven firehose
instability, the mirror instability passes through
$(\omega,\gamma)=(0,0)$, but unlike the \Alfven firehose, the related
stable mode is non-propagating, meaning that the return to a
marginally stable state of the plasma will not result in the
production of propagating waves.  For most values of $T_{\perp
  p}/T_{\parallel p}$, the Alfv\'en, fast, and slow modes are
essentially unchanged, with the exception of very large temperature
anisotropies.  For $T_{\perp p}/T_{\parallel p}\sim 10$, the proton
cyclotron instability goes unstable, even at the oblique wavevector
angles relevant to the mirror instability, as seen in the upper right
panel of Figure~\ref{fig:parameter}.

The bottom row of Figure~\ref{fig:instable_map} illustrates the
macroscopic nature of the mirror instability by plotting the growth
rate $\gamma/\Omega_p>0$ as a function of wavevector for unstable
modes with $\beta_{\parallel p}=1.0$ and $T_{\perp p}/T_{\parallel
  p}=2.0$ (left column) and $4.0$ (center). In both cases, the
unstable modes fill a broad region with $k_\parallel d_p \le k_\perp
\rho_p$ and $ k_\perp \rho_p \lesssim 1$.  Within the unstable
region, the growth rate $\gamma/\Omega_p$ increases linearly with
$k_\parallel d_p$, so the most rapid growth occurs for the largest
unstable values of $k_\parallel d_p$. Increased anisotropy leads to an
increase in the growth rate and increases the area of the unstable
wavevector region slightly, though the unstable region remains bounded by
$k_\perp \rho_p \lesssim 1$.

The fourth column of Figure~\ref{fig:aleph_scan} illustrates the
$(\beta_{\parallel p},T_{\perp p}/T_{\parallel p})$ frequency and
eigenfunction dependence of the entropy mode for $k_\perp \rho_p =
0.6$ and $k_\parallel d_p = 0.1$.  The entropy modes are
non-propagating, with zero real frequency $\omega =0$, for all \Bpar
and $T_{\perp p}/T_{\parallel p}$ (top row).  The damping rate (second
row) has a slight dependence on $\beta_{\parallel p}$, with larger
\Bpar generally leading to larger damping or growth rates. For a fixed
wavevector, the anisotropy value $T_{\perp p}/T_{\parallel p}$ at
marginal stability increases with decreasing $\beta_{\parallel p}$.
The magnetic eigenfunctions for the mirror-instability-driven entropy
modes are generally dominated by $\delta B_z$, with a subdominant
contribution from $\delta B_x$ which increases with increasing
$\beta_{\parallel p}$.  The mirror-instability-driven entropy mode is
the most compressible of the four modes considered in this
investigation, with larger density fluctuations for smaller \Bpar and
larger $T_{\perp p}/T_{\parallel p}$.

\section{Impact of Proton Temperature Anisotropy on the Large-Scale Cascade}
\label{sec:turb}

Now that we have established the properties of the four proton
temperature anisotropy instabilities, we consider how the proton
temperature anisotropy impacts the nonlinear dynamics of the \Alfvenic
fluctuations underlying the large-scale cascade.

Early research on incompressible MHD turbulence in the
1960s\cite{Iroshnikov:1963,Kraichnan:1965} emphasized the wave-like
nature of turbulent plasma motions, suggesting that nonlinear
interactions between counterpropagating \Alfven waves---or \Alfven
wave collisions---mediate the turbulent cascade of energy from large
to small scales. Following significant previous studies on weak
incompressible MHD
turbulence,\cite{Sridhar:1994,Montgomery:1995,Ng:1996,Galtier:2000}
the nonlinear energy transfer in \Alfven wave collisions has recently
been solved analytically in the weakly nonlinear limit,
\cite{Howes:2013a} confirmed numerically with gyrokinetic
simulations,\cite{Nielson:2013a} and verified experimentally in the
laboratory,\cite{Howes:2012b,Howes:2013b,Drake:2013} establishing
\Alfven wave collisions as the fundamental building block of
astrophysical plasma turbulence.
A discussion of the important role of linear wave physics in 
the context of strong turbulence can be found in 
Howes, Klein, \& TenBarge 2014\cite{Howes:2014b}
and Howes 2015.\cite{Howes:2014e}

The physics of \Alfven wave collisions in the inertial range
is most clearly illustrated 
in the context of the incompressible MHD equations expressed in the 
symmetrized Elsasser form,\cite{Elsasser:1950}
\begin{equation}
\frac{\partial \V{z}^{\pm}}{\partial t} 
\mp \V{v}_A \cdot \nabla \V{z}^{\pm} 
=-  \V{z}^{\mp}\cdot \nabla \V{z}^{\pm} -\nabla P/\rho_0,
\label{eq:elsasserpm}
\end{equation}
and $\nabla\cdot \V{z}^{\pm}=0$.  Here $\V{v}_A =\V{B}_0/\sqrt{4
  \pi\rho_0}$ is the \Alfven velocity due to the equilibrium field
$\V{B}_0=B_0 \hat{\V{z}}$ where $\V{B}=\V{B}_0+ \delta \V{B} $, $P$ is total
pressure (thermal plus magnetic), $\rho_0$ is mass density, and
$\V{z}^{\pm} = \V{u} \pm \delta \V{B}/\sqrt{4 \pi \rho_0}$ are the
Elsasser fields which represent waves that propagate up or down the
mean magnetic field. The $\V{z}^{\mp}\cdot \nabla \V{z}^{\pm} $ term
governs the nonlinear interactions between counterpropagating \Alfven
waves, or \Alfven wave collisions.

The mathematical form of the nonlinear term, $\V{z}^{\mp}\cdot \nabla
\V{z}^{\pm} $, determines important properties that govern the
turbulent cascade of energy from large to small scales. The first, and
most fundamental, property is that only counterpropagating waves
interact
nonlinearly,\cite{Iroshnikov:1963,Kraichnan:1965,Goldreich:1995,Howes:2013a}
a property that is possible because \Alfven waves (and pseudo-\Alfven
waves in the case of incompressible MHD) are not dispersive.  If not,
waves traveling in the same direction could catch up with each other
and interact nonlinearly. Second, the property that only
counterpropagating waves interact nonlinearly fundamentally leads to
an anisotropic cascade of energy in plasma
turbulence,\cite{Howes:2014e} in which energy is preferentially
transferred to small perpendicular scales, leading to small-scale
turbulent fluctuations with an anisotropy $k_\perp \gg k_\parallel$.
In this anisotropic limit, the \Alfven waves dominate the nonlinear
interactions,\cite{Maron:2001,Schekochihin:2009,Howes:2013a,Howes:2014c}
with compressible fluctuations relegated to a subdominant role in the
nonlinear energy transfer. Third, in order for the nonlinearity to be
nonzero, the vector nature of the nonlinear term requires that two
counterpropagating \Alfven waves, each of which is linearly polarized
in the MHD limit $k \rho_p \ll 1$, not be polarized in the same
plane.\citep{Howes:2013a} Therefore, to understand how the proton
temperature anisotropy impacts the large-scale turbulent cascade, it is
important to explore how \tempAni $\ne 1$ affects these two important
properties of the anisotropic \Alfven waves that control the turbulent
cascade: (1) the nondispersive nature of the \Alfven waves, and (2)
the polarization of the \Alfven wave eigenfunction.

The large-scale cascade is dominated by \Alfvenic fluctuations in an
anisotropic region of wavevector space with $k_\perp \gg
k_\parallel$. This sense of anisotropy is well-supported by
multi-spacecraft measurements of turbulence near the ion kinetic
scales in the solar
wind.\cite{Sahraoui:2010b,Narita:2011,Roberts:2013} Modern scaling
theories of \Alfvenic turbulence suggest that a critical balance
between linear and nonlinear timescales leads to turbulent
fluctuations that obey a scale-dependent wavevector
anisotropy.\cite{Goldreich:1995,Boldyrev:2006} To explore the effect
of the proton temperature anisotropy \tempAni on these anisotropic
\Alfven waves, we solve for their properties along an idealized line
of critical balance given
by\cite{Howes:2008b,Howes:2011b,TenBarge:2012a}
\begin{equation}
k_\parallel \rho_p = (k_0 \rho_p)^{1/3} 
\left[\frac{(k_\perp \rho_p)^{2/3} + (k_\perp \rho_p) ^{7/3}}{1+(k_\perp \rho_p)^2}\right],
\label{eq:critbal}
\end{equation}
where $k_0 \rho_p = 10^{-3}$ is the isotropic driving scale (the outer
scale of the turbulent inertial range), at which fluctuations are
assumed to be isotropic with $k_\perp = k_\parallel$. This critical
balance relation represents the approximate upper boundary (in
$k_\parallel$) in wavevector space of the power in the anisotropic
turbulent fluctuations.\cite{Howes:2008b,Howes:2011b} Limits of this
idealized relation lead to a wavevector anisotropy scaling as
$k_\parallel \propto k_\perp^{2/3}$ in the MHD inertial range
($k_\perp \rho_p \ll 1)$ and $k_\parallel \propto k_\perp^{1/3}$ in
the kinetic dissipation range ($k_\perp \rho_p \gtrsim
1$).\cite{Howes:2015book} The wavevector scalings in those two limits
are supported by MHD simulations of \Alfven wave
turbulence\cite{Cho:2000,Maron:2001} and gyrokinetic simulations of
kinetic \Alfven wave turbulence.\cite{TenBarge:2012a}

\begin{figure*}[t]
\begin{center}
\includegraphics[width=15.00cm,viewport=10 5 425 430, clip=true]
{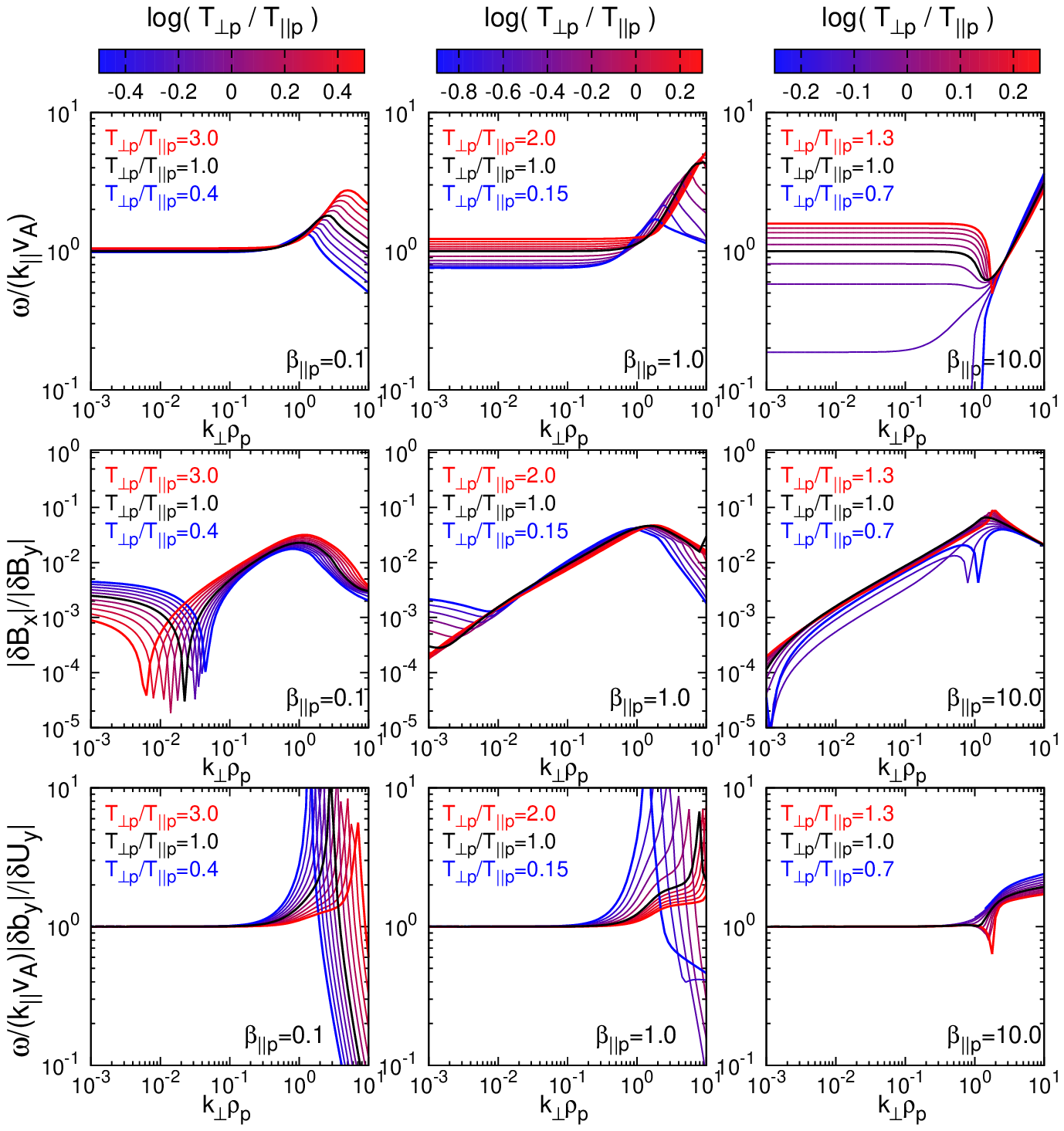}\\
\caption[] { Plots of $\omega/k_\parallel v_{A}$ (top row),
  $|\delta B_x|/|\delta B_y|$ (center), and $(\omega/k_\parallel
  v_{A}) \delta b_y/\delta U_y$ (bottom) for \Alfven waves along
  critical balance, Equation~\ref{eq:critbal}.  
  Here $\delta b_y = \delta B_y/\sqrt{4 \pi n_p m_p}.$
  Color indicates temperature anisotropy, with
  \tempAni $>1$ $(<1)$ shown in red (blue) and \tempAni $=1$ plotted
  in black.  The columns are organized by $\beta_{\parallel p}$, with
  \Bpar $=0.1$ (left column), $1.0$ (center), and $10.0$ (right).
  Different limits on \tempAni are chosen for each \Bpar to
  approximate solar wind observations. }
\label{fig:turb_scan}
\end{center}
\end{figure*}

In Figure~\ref{fig:turb_scan}, we plot the real frequency as well as
two metrics of the eigenfunction polarization for \Alfven waves 
for the Vlasov-Maxwell system along
this idealized critical-balance line. The columns are organized by
$\beta_{\parallel p}$, with \Bpar $=0.1$ (left column), \Bpar $=1.0$
(center), and \Bpar $=10.0$ (right). We explore the range of proton
temperature anisotropies that are observed in the solar
wind\cite{Bale:2009} at each value of $\beta_{\parallel p}$, so each
column has different values for $(T_{\perp p }/T_{\parallel
  p})_{\mbox{min}}$ and $(T_{\perp p }/T_{\parallel p})_{\mbox{max}}$.
The temperature anisotropy is indicated by the color of each curve,
with the colorbar displaying the variation of colors from the minimum
to the maximum temperature anisotropy for that column. The minimum
proton temperature anisotropy $(T_{\perp p }/T_{\parallel
  p})_{\mbox{min}}$ is blue, an isotropic proton temperature \tempAni
$=1$ is black, and the maximum $(T_{\perp p }/T_{\parallel
  p})_{\mbox{max}}$ is red.

\Alfven waves propagate along the local mean magnetic field $\V{B}_0$
at the parallel group velocity $v_{g \parallel} = \partial
\omega/\partial k_\parallel$. If this group velocity is independent of
the wavevector, then the \Alfven waves are nondispersive and an
arbitrary \Alfven wavepacket will propagate without distortion along
the magnetic field at the parallel group velocity. In the top row of
Figure~\ref{fig:turb_scan}, the normalized frequency
$\omega/k_\parallel v_{ A}$ is plotted vs.~normalized perpendicular
wavenumber $k_\perp \rho_p$ along the path of critical balance given
by \eqref{eq:critbal}. For each value of $\beta_{\parallel p}$, we see
that the \Alfven waves are indeed nondispersive (the normalized curve
$\omega/k_\parallel v_{ A}$ is independent of $k_\perp$, and therefore
$v_{g \parallel}$ is constant), except for $k_\perp \rho_p \gtrsim 1$
where finite proton Larmor radius effects lead to a transition to
dispersive kinetic \Alfven waves.

For an isotropic proton temperature \tempAni $= 1$, the \Alfven waves
propagate at the \Alfven velocity $v_A$ for any value of
$\beta_{\parallel p}$ (black). For an anisotropic proton temperature
\tempAni $\neq 1$, the parallel group velocity depends on the value of
\tempAni (particularly for $\beta_{\parallel p}\ge 1$), but for any
particular choice of plasma parameters $(\beta_{\parallel p}, T_{\perp
  p}/T_{\parallel p})$, the \Alfven wave remains nondispersive in the
inertial range $k_\perp \rho_p \ll 1$.  In fact, the dispersion
relation is well modeled by the double adiabatic CGL
result,\cite{Chew:1956,Krall:1973} $\omega = k_\parallel v_{A}\sqrt{1
  + \beta_\parallel (T_\perp/T_\parallel-1)/2}$. Therefore, the
nondispersive nature of \Alfven waves is insensitive to the proton
temperature anisotropy, and so we expect that the property that only
counterpropagating waves interact nonlinearly persists for \tempAni
$\neq 1$. Of course, when \Alfven waves become unstable to the
macroscopic CGL firehose instability, they become non-propagating with
$\omega = 0$, as can be seen for the two lowest values of $T_{\perp
  p}/T_{\parallel p}$ (blue) in the \Bpar $=10$ column of
Figure~\ref{fig:turb_scan}.  However, a nearly negligible fraction of
observed solar wind intervals have parameters unstable to the CGL
firehose,\cite{Hellinger:2006,Bale:2009,Maruca:2011} reducing the
significance of these non-propagating modes for the large-scale
turbulent cascade. It is worthwhile emphasizing, also, that the
parallel firehose and \Alfven firehose velocity-space instabilities
are unstable in regions of wavevector space (see
Figure~\ref{fig:instable_map}) that do not overlap the anisotropic
region of wavevector space inhabited by the \Alfvenic fluctuations
associated with the large-scale cascade, given by modes with parallel
wavenumbers below the critical balance line \eqref{eq:critbal}.

Next, we investigate whether an anisotropic proton temperature alters
the polarization of the anisotropic \Alfven waves associated with the
large-scale cascade. For an isotropic temperature \tempAni $= 1$,
\Alfven waves are linearly polarized, with $|\delta B_y| \gg |\delta
B_x| \sim |\delta B_z|$. To determine any anisotropy-induced
deviations from linear polarization, we plot $|\delta B_x|/|\delta
B_y|$ in the second row of Figure~\ref{fig:turb_scan} for the same
range of plasma parameters $(\beta_{\parallel p}, T_{\perp
  p}/T_{\parallel p})$.  The \Alfven wave remains linearly polarized,
with $|\delta B_y| \gg |\delta B_x|$, with little change from the
isotropic temperature case \tempAni $= 1$ (black) for all values in
the observed range of $(\beta_{\parallel p}, T_{\perp p}/T_{\parallel
  p})$.  This result holds even for the modes unstable to the CGL
firehose in the \Bpar$=10$ case. With no changes in the polarization
of the \Alfven waves, we expect no significant modifications of the
nonlinearity responsible for the large-scale turbulent cascade due to
proton temperature anisotropy.

Finally, for an isotropic temperature \tempAni $= 1$, the
eigenfunction relation for \Alfven waves is given by $\delta B_y/B_0 =
\pm \delta u_y/v_A$, where the sign dictates the direction of
propagation of the \Alfven wave along the mean magnetic field
$\V{B}_0$. This property has been used to identify large-scale \Alfven
waves in the solar wind\cite{Belcher:1971} and is the physical basis
enabling upward and downward propagating \Alfven waves to be described
by the Elsasser variables, $\V{z}^{\pm}/v_A = \V{u}/v_A \pm \delta
\V{B}/B_0$.  Using the double adiabatic CGL dispersion relation for
MHD \Alfven waves, we obtain an eigenvalue relation generalized to
account for temperature anisotropy, $\omega/(k_\parallel v_{A}) \delta
B_y/B_0 = \pm \delta u_y/v_A$. This simple result enables one to
construct generalized Elsasser variables $\V{z}_G^{\pm}$ for the case
of anisotropic temperature, $\V{z}_G^{\pm}/v_A = \V{u}/v_A \pm
\omega/(k_\parallel v_{A}) \delta\V{B}/B_0$. We test this
generalization of the Elsasser variables by plotting
$\omega/(k_\parallel v_{A}) |\delta B_y/B_0|/|\delta u_y/v_A|$
vs.~$k_\perp \rho_p$ in the third row of Figure~\ref{fig:turb_scan},
demonstrating clearly the validity of this generalization of the
Elsasser variables for \Alfven waves throughout the inertial range
$k_\perp \rho_p \ll 1$ for any choice of plasma parameters
$(\beta_{\parallel p}, T_{\perp p}/T_{\parallel p})$.

In summary, we find that neither the nondispersive nature nor the
polarization of the \Alfven waves that constitute the large-scale
turbulent cascade are altered by the proton temperature anisotropy.
Therefore, we conclude that the physics of the large-scale cascade in
the inertial range is insensitive to the proton temperature anisotropy
over the range of values observed in the solar wind.  Consequently, we
expect that studies of \Alfven wave nonlinear turbulent interactions
using an isotropic proton velocity distribution will still accurately
describe turbulence in the inertial range properly, even for plasmas
with \tempAni $\neq 1$.

The empirical prediction presented in this section, that the
turbulent dynamics of the large-scale cascade is not significantly
altered by proton temperature anisotropy, is supported by the results
of a recent theoretical treatment of kinetic turbulence in the
inertial range for plasmas with anisotropic temperature distributions
or relative drift among ion species.\cite{Kunz:2015}  In that study,
the main physical features of plasma turbulence in the inertial range
persist for non-Maxwellian distributions: the Alfvenic and compressive
fluctuations are decoupled, with the latter passively advected by the
former; and the Alfvenic cascade remains essentially fluid with
dynamics governed by the equations of reduced MHD modified to account
for changes in the Alfven speed due to pressure anisotropy and species
drifts.


\section{Energy Injection by Proton Temperature Anisotropy Instabilities}
\label{sec:injection}

Although the proton temperature anisotropy does not appear to alter
significantly the nature of the turbulent nonlinear interactions
involved in the anisotropic cascade of energy from large to small
scales, kinetic instabilities can drive electromagnetic fluctuations
in the plasma, thereby directly injecting energy into fluctuations at
kinetic scales.  Here we address two key questions about these
instability-driven fluctuations: (1) What are the observable
signatures of the instability-driven fluctuations, especially
considering that spacecraft measurements are made both in a moving
frame of reference and in the presence of the large-scale turbulent
cascade?; and (2) How do these instability-driven fluctuations
interact nonlinearly with each other and with the fluctuations of the
large-scale cascade?

\subsection{Observable Signatures}
\label{sec:obs}

Let us consider first the instability-driven fluctuations in the frame
of the solar wind plasma, a frame of reference generally moving at a
super-\Alfvenic velocity with respect to the spacecraft frame in which
measurements are made.\cite{Tu:1995,Bruno:2005} The two parallel
proton temperature anisotropy instabilities, the proton cyclotron and
parallel firehose instabilities, generate propagating proton cyclotron
and whistler waves with nonzero frequencies (see lower panels in
Figure~\ref{fig:parameter}) in the wavevector region with $0.1 \le
k_\parallel d_p \le 1$ and $k_\perp \ll k_\parallel$ (see top two rows
of Figure~\ref{fig:instable_map}).  These fluctuations may serve to
transport thermal energy from one spatial region of the solar wind to
another via the Poynting flux of the unstable waves.  On the other
hand, the two oblique instabilities, the \Alfven firehose and mirror
instabilities, generate non-propagating fluctuations with $\omega=0$
in the plasma frame (see upper panels of Figure~\ref{fig:parameter}).
The \Alfven firehose instability drives fluctuations with peak growth
rates in the wavevector region with $0.1 \le k_\parallel d_p \le 0.4$
and $0.1 \le k_\perp \rho_p \le 1$ (see third row of
Figure~\ref{fig:instable_map}). The mirror instability drives
fluctuations over a broad wavevector region with $k_\parallel <
k_\perp$ and $k_\perp \rho_p \lesssim 1$, but its peak growth rates
(which increase linearly with $k_\parallel$) are restricted to the
smallest scales, $0.1 \le k_\perp \rho_p \le 1$ (see bottom row of
Figure~\ref{fig:instable_map}).  We expect that the more slowly
growing mirror modes at $k_\perp \rho_p \lesssim 0.1$ will not be
observable because the fluctuations of the large-scale cascade
occupying the same region of wavevector space, which increase in
amplitude with decreasing wavenumber, will be significantly larger and
therefore dominate measurements at these low perpendicular
wavenumbers.

Next we consider the signature of these instability-driven
fluctuations as measured in the spacecraft frame as the solar wind
plasma flows past at a super-\Alfvenic velocity.  In general, for a
fluctuation with wavevector $\V{k}$ and plasma-frame frequency
$\omega$ flowing past the sampling spacecraft at solar wind velocity
$\V{v}_{\mbox{sw}}$, the spacecraft-frame frequency is given by
$\omega_{sc} = \omega + \V{k}\cdot \V{v}_{sw}$, the sum of the
plasma-frame frequency plus a Doppler-shifted spatial
variation.\cite{Howes:2014a} For a sufficiently fast flow speed, the
plasma-frame frequency term is negligible compared to the Doppler
shift term, so $\omega_{sc} \simeq \V{k}\cdot \V{v}_{sw}$, an
approximation known as the Taylor hypothesis,\cite{Taylor:1938} widely
employed in the solar wind since the solar wind speed is typically
$v_{sw} \sim 10 v_A$. The condition for the Taylor hypothesis to be
valid for both \Alfven and whistler waves\cite{Howes:2014a} is
$(v_{sw}/v_A) \cos \theta_{kv} \gg k_\parallel d_p$, where $\V{k}\cdot
\V{v}_{sw}= k v_{sw} \cos \theta_{kv}$. Since all of these
instability-driven modes satisfy $k_\parallel d_p \lesssim 1$, we may
safely adopt the Taylor hypothesis. An important point to emphasize
here is that for choices of angle $ \theta_{kv}$ such that $\cos
\theta_{kv}\ll 1$ (a condition that would violate the Taylor
hypothesis), the spacecraft-frame frequency of the instability-driven
fluctuation is downshifted sufficiently that it will be unmeasurable
in the presence of the large-scale cascade.

If we express the wavevector of a particular instability-driven mode
as $\V{k}= k_\parallel \hat{\V{b}} +\V{k}_\perp$, where the direction
of the local mean magnetic field is given by the unit vector
$\hat{\V{b}} = \V{B}_0/B_0$, the spacecraft-frame frequency can be
written
\begin{equation}
\omega_{sc} \simeq k_\parallel v_{sw} \cos \theta_{vB} + \V{k}_\perp\cdot \V{v}_{sw},
\end{equation}
where the angle $ \theta_{vB}$ is given by $\V{v}_{sw} \cdot \V{B}_0 =
\cos \theta_{vB}$.  For the proton cyclotron and parallel firehose
instabilities, the wavevectors of unstable modes satisfy $k_\perp \ll
k_\parallel$, so this expression reduces to $\omega_{sc} \simeq
k_\parallel v_{sw} \cos \theta_{vB}$.  Note that the solar wind speed
$v_{sw}$ and the angle between the solar wind flow and the local mean
magnetic field $\theta_{vB}$ are both directly measurable from
spacecraft measurements, so this calculation establishes a direct
relation between the spacecraft-frame frequency and parallel
wavenumber for the parallel modes arising from these two
instabilities.

We must also consider whether these fluctuations can be measured in
the presence of the broadband fluctuations due to the large-scale
cascade.  The amplitude of the background turbulent fluctuations
decreases with increasing frequency, so these parallel
instability-driven waves are most likely to be measurable at their
maximum spacecraft-frame frequency $\omega_{sc} \simeq k_\parallel
v_{sw}$, occurring when $\theta_{vB}\rightarrow 0$, or physically when
the direction of the local mean magnetic field aligns with the solar
wind flow.  Such alignment between the magnetic field and the solar
wind flow, $\theta_{vB}\rightarrow 0$, has previously been exploited
to study the variation of solar wind turbulence parallel to the
magnetic field in a number of observational
studies.\cite{Horbury:2008,Podesta:2009,Wicks:2010,He:2011a,Podesta:2011a,Forman:2011,Chen:2012b,Wicks:2012}
We can use this maximum-frequency relation 
and the wavevector region of unstable modes to
predict the linear spacecraft-frame frequency for the fluctuations
driven by the proton cyclotron and parallel firehose instabilities,
\begin{equation}
0.02 \frac{v_{sw}}{d_p} \le f_{sc} \le 0.2 \frac{v_{sw}}{d_p}, 
\label{eq:freq_par}
\end{equation}
where $\omega_{sc}=2 \pi f_{sc}$.

Although the \Alfven firehose and mirror instabilities generate
unstable non-propagating fluctuations with real frequency $\omega=0$
in the plasma frame, in the spacecraft frame these fluctuations have
$\omega_{sc} \ne 0$ due to the Doppler shift term. Thus, it is
difficult, from single-point spacecraft measurements alone, to
distinguish these non-propagating instability-driven fluctuations from
propagating waves in the solar wind.  The Taylor hypothesis is, of
course, trivially satisfied for these unstable modes since $\omega=0$,
yielding the equality $\omega_{sc} = \V{k}\cdot \V{v}_{sw}$. Again, in
the presence of the large-scale cascade, these instability-driven
fluctuations are most easily measured at the maximum spacecraft-frame
frequency $\omega_{sc} \simeq k v_{sw}$ occurring when
$\theta_{kv}\rightarrow 0$.  Although $\theta_{kv}$ is not an
observationally accessible quantity using single-point spacecraft
measurements, this simple expression provides a valuable estimate for
the predicted spacecraft-frame frequency of potentially observable
fluctuations driven by the \Alfven firehose and mirror instabilities.
Since these modes have peak growth rates with $0.3 k_\perp \rho_p
\lesssim k_\parallel d_p \lesssim k_\perp \rho_p$ for
$\beta_{\parallel p}\sim 1$ (see bottom two rows of
Figure~\ref{fig:instable_map}), the magnitude of the wavevector can be
estimated by $k = (k_\parallel^2 + k_\perp^2)^{1/2} \sim k_\perp$,
yielding a simplified prediction for the linear spacecraft-frame
frequency of fluctuations driven by the \Alfven firehose and mirror
instabilities,
\begin{equation}
0.02 \frac{v_{sw}}{\rho_p} \le f_{sc} \le 0.2 \frac{v_{sw}}{\rho_p}.
\label{eq:freq_perp}
\end{equation}
Note that this expression is sensitive to the minimum value of
$\theta_{kv}$ at which there is energy from instability-driven
fluctuations; for example, if the solar wind flow is instantaneously
along the local mean magnetic field ($\theta_{vB}=0$), then the
unstable regions of wavevector space in the bottom row of
Figure~\ref{fig:instable_map} would yield $\theta_{kv}\ge \pi/4$,
reducing the measured $\omega_{sc}$ by a factor $\cos \theta_{kv}
=0.71$ from this simple estimate.

From this analysis, a key qualitative difference between the turbulent
fluctuations of the large-scale cascade and the instability-driven
fluctuations with solar wind relevant plasma parameters
is their extent in frequency. The cascade of turbulent
fluctuation energy from large to small scales is characterized by a
broadband frequency spectrum of turbulent fluctuations. In contrast,
the instability-driven fluctuations are likely to exhibit a narrowband
frequency spectrum within the ranges given by \eqref{eq:freq_par} and 
\eqref{eq:freq_perp}. 

Of course, whether the fluctuations driven by any of the four proton
temperature anisotropy instabilities is observable in the presence of
the large-scale cascade depends on the amplitudes of both the
instability-driven fluctuations and the turbulent fluctuations
comprising the large-scale cascade. The amplitude of the
instability-driven fluctuations depends on the nonlinear saturation
mechanism for each instability as well as the large-scale dynamics
pushing the plasma into an unstable region of plasma parameter space,
both topics beyond the scope of this study.

Even without a prediction for the saturated amplitudes of the
instability-driven modes, a definitive identification of these modes
may be possible using the properties of these modes that
observationally distinguish them from the dominant \Alfvenic
fluctuations and subdominant compressible fluctuations of the
large-scale cascade. A careful comparison of the correlations between
two components of the measurements with that of the linear
eigenfunctions for each instability-driven mode can be used to make
such an identification.\cite{Howes:2014b} There is a well-established
precedent for such identifications using various polarizations and
correlations \cite{Gary:1992,Gary:1993,Song:1994,Lacombe:1995} in the
magnetosphere\cite{Denton:1995,Schwartz:1996} and in the bulk of the
solar
wind.\cite{Leamon:1998b,Howes:2012a,Klein:2012,Salem:2012,TenBarge:2012b,Chen:2013a,Roberts:2013,Klein:2014a,Lacombe:2014}
An example of discriminating properties from the eigenfunctions of the
instability-driven waves in Figure~\ref{fig:aleph_scan} include the
fact that the proton cyclotron waves driven by the proton cyclotron
instability have left-handed circular magnetic polarization and are
largely incompressible (first column), while the whistler waves driven
by the parallel firehose instability have right-handed circular
magnetic polarization and are also largely incompressible (second
column). It is worthwhile noting that the properties of both the
proton cyclotron waves and the whistler waves driven by these two
instabilities are largely insensitive to both the parallel proton
plasma beta $\beta_{\parallel p}$ and the proton temperature
anisotropy $ T_{\perp p}/T_{\parallel p}$, potentially making
eigenfunction tests of these modes particularly robust to plasma
parameter variations. The properties of the fluctuations arising from
the \Alfven firehose and mirror instabilities have substantially more
variation with changes in the plasma parameters, but it may be
possible to devise discriminating tests for these modes as well.

We conclude this section with a brief review of potential observations
of fluctuations caused by proton temperature anisotropy instabilities
in the literature.  A wavelet analysis of measurements of the solar
wind magnetic energy spectrum as a function of $\theta_{vB}$ within
high-speed streams in the ecliptic solar wind using the Stereo
spacecraft, presented in Podesta~2009,\cite{Podesta:2009} shows a bump
in the nearly parallel bin $ \theta_{vB}=3^\circ$ at $f_{sc} \simeq
0.4$~Hz that may be a signature of fluctuations driven by one of these
instabilities, although the paper does not summarize the plasma
parameters of the measured intervals to test whether they agree with
the quantitative predictions of $f_{sc}$ presented here.  The same
wavelet analysis technique applied to Ulysses measurements from the
fast polar solar wind in Wicks \emph{et al.} 2010\cite{Wicks:2010}
found a bump in the parallel spectrum, binned over $0^\circ \le
\theta_{vB} < 10^\circ$, at a value of $k\rho_p \sim 0.6$ that may
also be due to instability-driven fluctuations. Measurements of the
reduced fluctuating magnetic
helicity\cite{Matthaeus:1982b,Goldstein:1994,Leamon:1998b,Hamilton:2008,Howes:2010a,Klein:2014a}
sorted by period $T$ and angle $\theta_{vB}$ find a broad
perpendicular signature of positive magnetic helicity and a narrow
parallel signature of negative magnetic
helicity.\cite{Podesta:2011a,He:2011a} The broad perpendicular
signature was interpreted to be due to an anisotropic distribution of
KAWs with $k_\perp \gg k_\parallel$ associated with the large-scale
cascade,\cite{He:2011a,Podesta:2011a} while the parallel signature was
proposed to arise from proton cyclotron waves propagating anti-sunward
along the magnetic field \cite{He:2011a,Podesta:2011a} driven by the
proton cyclotron instability or from whistler waves propagating
sunward \cite{Podesta:2011a,Podesta:2011b} driven by the parallel
firehose instability. A comparison of the magnetic helicity from these
spacecraft measurements to that produced using the synthetic
spacecraft data method\cite{Klein:2012} has constrained the energy
content of the instability-driven parallel modes to be around 5\% of
the energy in the large-scale cascade over the narrow frequency band
of the parallel magnetic helicity signature.\cite{Klein:2014a} Another
example, although driven by a kinetic electron rather than a kinetic
proton instability, is the recent study by Lacombe \emph{et al.}
2014\cite{Lacombe:2014} which presents evidence of power in
intermittent whistler waves at frequencies $10$~Hz~$\lesssim f_{sc}
\lesssim 100$~Hz, interpreted to be driven by the whistler heat flux
instability.\cite{Lacombe:2014} These studies provide significant
motivation for more thorough investigations to identify fluctuations
driven by proton temperature anisotropy instabilities in the solar
wind and to constrain the energy content of these fluctuations and their
effect on the thermodynamic evolution of the solar wind.

An important open question is how the presence of pre-existing
turbulence impacts the linear frequencies and growth rates of kinetic
temperature anisotropy instabilities. The treatment presented here
assumes that, to lowest order, linear theory predictions for the
growth rates in a quiescent medium give reasonable values for
estimation. The observational evidence for modes driven by parallel
instabilities, reviewed above, suggests that these instabilities do
indeed operate within the turbulent solar wind plasma.

\subsection{Nonlinear Interactions}
\label{sec:nl}

Next we consider how the fluctuations driven by proton temperature
anisotropy instabilities interact nonlinearly with each other and with
the anisotropic fluctuations of the large-scale turbulent cascade.

There are, in fact, several lines of reasoning that suggest that the
instability-driven fluctuations experience negligible nonlinear
interactions with each other. First, the parallel signature seen in
magnetic helicity measurements\cite{He:2011a,Podesta:2011a} is found
to have a normalized value near 1.0, suggesting that the
instability-driven waves, if they are responsible for the signature,
propagate almost entirely in one direction along the magnetic field,
either anti-sunward for proton cyclotron waves or sunward for whistler
waves.\cite{He:2011a,Podesta:2011a,Podesta:2011b,Klein:2014a} For the
non-dispersive \Alfven waves dominating the inertial range of solar
wind turbulence, the nonlinear interactions that mediate the turbulent
cascade of energy occur only between counterpropagating
waves.\cite{Iroshnikov:1963,Kraichnan:1965,Sridhar:1994,Montgomery:1995,Ng:1996,Galtier:2000,Howes:2012b,Drake:2013,Howes:2013a,Howes:2013b,Nielson:2013a}
Although this property does not strictly hold true for dispersive
waves (since waves with higher perpendicular wavenumbers can overtake
waves with lower perpendicular wavenumbers), any nonlinear
interactions that do occur may be weaker by virtue of the longer
timescales for waves to overtake each other. Second, the energy
content of the instability-driven waves has been found in a 
recent study to be only 5\% of the large-scale
cascade fluctuation energy over the same spacecraft-frame frequency
band.\cite{Klein:2014a} Therefore, these unstable waves either have a
small amplitude or are spatially intermittent.  Small amplitude
implies a weaker nonlinearity, and spatial intermittency implies lower
probability that two wavepackets will collide and interact
nonlinearly; either way, the nonlinear interactions are likely to be
weaker than those associated with the large-scale cascade.  Finally,
the potential observations of instability-driven waves reviewed above
are all qualitatively narrowband in spacecraft-frame frequency,
whereas a typical characteristic of turbulence with strong
nonlinearities is a broadband spectrum of fluctuations, suggesting
that these instability-driven waves do not undergo a significant
turbulent cascade.

Finally, we consider nonlinear interactions of the instability-driven
fluctuations with the turbulent fluctuations of the large-scale
cascade. The first key point is that, with the exception of the mirror
instability, the unstable fluctuations generated by the proton
temperature anisotropy instabilities arise in a region of wavevector
space (generally $0.1 \le k_\parallel d_p \le 1$ and $k_\perp \lesssim
k_\parallel$) that does not overlap with the anisotropic fluctuations
of the large-scale cascade in the region $k_\perp \gg k_\parallel$.
Even for the mirror instability, the peak growth rates occur at
moderately oblique wavevectors with $k_\perp \rho_p \sim 3 k_\parallel
d_p$, wavevectors adjacent to, but not generally within, the region
$k_\perp \gg k_\parallel$ of the large-scale cascade.

\begin{figure}[t]
\begin{center}
\includegraphics[width=7.00cm,viewport=25 22 127 115, clip=true]
{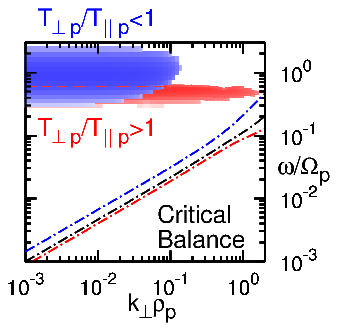}\\
\caption[]
{
Dispersion relations $\omega(k_\perp \rho_p)/\Omega_p$ for
stable \Alfven waves along the line of critical balance (dash-dotted lines)
and for unstable proton cyclotron and parallel firehose modes with 
$\gamma/\Omega_p>10^{-4}$, respectively illustrated by the
partially overlapping red and blue shaded regions.
Plasma parameters are set to  
\tempAni $=2.68$ (red), \tempAni $=1.00$ (black), and \tempAni $=0.27$ (blue)
with \Bpar $=1.7$.
}
\label{fig:impedence}
\end{center}
\end{figure}

For the propagating waves driven by the proton cyclotron and parallel
firehose instabilities, this mismatch in wavevector space leads to
significantly higher plasma-frame frequencies for the
instability-driven waves than the anisotropic, low-frequency
fluctuations of the large-scale cascade.  In
Figure~\ref{fig:impedence}, we plot the normalized plasma-frame
frequency $\omega/\Omega_p$ for both \Alfven waves along critical
balance (dash-dotted lines) and the unstable modes generated by the
parallel instabilities with $\gamma/\Omega_p > 10^{-3}$ (shaded region).
The proton temperature anisotropy is varied from \tempAni $=2.68$
(red) to \tempAni $=1.00$ (black) to \tempAni $=0.27$ (blue), with
\Bpar $=1.7$.  
This significant difference in plasma-frame
frequencies results in an effective impedance mismatch that is
expected to prevent significant nonlinear interactions between the
high-frequency unstable waves and the low-frequency turbulent
fluctuations. Physically, high-frequency fluctuations experience much
lower frequency fluctuations as a nearly static background variation,
while low-frequency fluctuations are only weakly affected by the rapid
oscillation of much higher frequency fluctuations. But, as these
instability-driven waves attempt to propagate along the tangled
magnetic field associated with the large-scale turbulent cascade,
wavepackets may be sheared out, effectively transferring their energy
to smaller perpendicular scales. Thus, the energy of the
instability-driven waves may be cascaded to smaller scales by the
fluctuations of the large-scale cascade, but the unstable waves are
not likely to alter significantly the turbulent nonlinear dynamics of
the large-scale cascade, especially if the instability-driven waves
have small amplitudes or are spatially intermittent.

Of course, the \Alfven firehose and mirror instabilities generate
non-propagating fluctuations with zero plasma-frame frequency,
$\omega=0$. These fluctuations will essentially form a static
background upon which the active large-scale turbulent cascade must
proceed.  It is possible that these modes can alter the nonlinear
energy transfer of the anisotropic \Alfvenic fluctuations, meriting
further consideration, although if the growth of these unstable
fluctuations ceases at small amplitude, their effect on the
large-scale cascade may be minimal. 
On the other hand, these static
background modes may indeed be cascaded to smaller scales by the
active \Alfvenic fluctuations of the large-scale cascade.
Future study into the effects of the turbulent cascade 
on the nonlinear saturation of these instabilities will be necessary
to determine if the associated unstable modes and nonlinear structures
can grow to sufficiently large amplitudes to affect the large-scale cascade.

\section{Conclusion}
\label{sec:conc}

In the weakly collisional solar wind plasma, non-Maxwellian particle
velocity distributions are routinely measured, but how these
non-Maxwellian distributions impact the physics of plasma turbulence
in the solar wind remains unanswered. Here we take a first step in
addressing this question by exploring in detail the effect of a
bi-Maxwellian proton temperature distribution on the plasma
turbulence.  Specifically, we aim to understand how the proton
temperature anisotropy affects the nonlinear dynamics of the
large-scale turbulent cascade and how the fluctuations driven by the
proton temperature anisotropy interact nonlinearly with each other and
with the fluctuations of the large-scale cascade.

For a plasma with a bi-Maxwellian proton distribution, there exist
at least four proton temperature anisotropy instabilities: the proton
cyclotron, parallel firehose, \Alfven firehose, and mirror
instabilities. We present a unified framework for these instabilities,
identifying the associated stable eigenmodes, highlighting the
unstable region of wavevector space, and presenting the properties of
the growing eigenfunctions.

We find that the proton temperature anisotropy significantly affects
neither the nondispersive nature nor the polarization of the \Alfven
waves that constitute the large-scale turbulent cascade, so we
conclude that the nonlinear interactions governing the turbulent
cascade of energy through the inertial range are insensitive to the
proton temperature anisotropy.  Consequently, we expect that studies
of the nonlinear dynamics of plasma turbulence assuming an isotropic
proton velocity distribution will still qualitatively and
quantitatively describe the lowest-order behavior of solar wind
turbulence in the inertial range, even for plasmas with \tempAni $\neq
1$.  The electromagnetic fluctuations driven by these temperature
anisotropy instabilities are expected to be observed within a
narrowband of linear spacecraft-frame frequencies $0.02 v_{sw}/r_p \le
f_{sc} \le 0.2 v_{sw}/r_p$, where $r_p$ is either the proton inertial
length or proton Larmor radius, and the properties of the unstable
linear eigenfunctions may be exploited to identify these
instability-driven fluctuations in spacecraft measurements.  We argue
that the instability-driven waves are unlikely to interact nonlinearly
with each other, and that, although their energy may be cascaded to
smaller scale by the turbulent fluctuations of the large-scale
cascade, it is unlikely that the instability-driven waves
significantly alter the nonlinear dynamics of the large-scale cascade.

These findings are fully consistent with a recent perspective on the
fluid and kinetic aspects of the weakly collisional plasma turbulence
in the solar wind\cite{Howes:2014e} as well as an analytic treatment
of temperature anisotropic inertial range kinetic turbulence.\cite{Kunz:2015}
In this picture, the nonlinear
interactions responsible for the turbulent cascade of energy and the
formation of current sheets in kinetic plasma turbulence are
essentially fluid in nature, and are not significantly impacted by the
typically non-Maxwellian form of the distribution functions. On the
other hand, the damping of the turbulent fluctuations of the
large-scale cascade and the injection of energy by kinetic
instabilities via collisionless wave-particle interactions are
essentially kinetic in nature, and thus may depend sensitively on the
non-Maxwellian form of the distribution functions. This viewpoint
strongly contradicts another recent study using a hybrid
Vlasov-Maxwell numerical approach which claims that the non-Maxwellian
features of the proton distribution function ``may be a key point for
understanding the complex nature of plasma turbulence'', and that
approaches employing Maxwellian or bi-Maxwellian distribution
functions have limited applicability for the study of solar wind
turbulence.\cite{Servidio:2014c} The study of kinetic turbulence, a
new frontier in heliophysics research, will clearly remain a hot topic
for the foreseeable future.

\begin{acknowledgments}
K.G.K. thanks Daniel Verscharen for useful discussions concerning
anisotropy instabilities.  This work was supported by NSF AGS-1331355,
NSF CAREER Award AGS-1054061 and NASA grant NNX10AC91G.
\end{acknowledgments}

\appendix

\section{Exceptional Points and Mode Identification}
\label{sec:EPs}

The delineation between distinct dispersion surfaces of the Vlasov-Maxwell system
breaks down in parameter regimes where the dispersion
relation contain branch point singularities,
known in this context as exceptional points.\citep{Kato:1966}
Following a path in parameter space 
which encircles such an exceptional point
results in a continuous transition between distinct solutions. 
For instance, consider following a particular solution 
$\omega_0$ along a path connecting 
$(A_1,B_1)$, $(A_2,B_1)$, $(A_2,B_2)$, 
$(A_1,B_2)$, and $(A_1,B_1)$, where $A$ and $B$
are parameters of the Vlasov-Maxwell system and $A_{1,2}$ and $B_{1,2}$
are chosen to enclose the exceptional point $(A^*,B^*)$;
$A^* \in (A_1,A_2)$ and $B^* \in (B_1,B_2)$.  The solution
$\omega_1$ found upon returning to 
$(A_1,B_1)$ will not be the same as the initial solution $\omega_0$.
To return to $\omega_0$ requires multiple revolutions around the exceptional point. 
In general, exceptional points signify that the dispersion surfaces
are not disjoint, but that the surfaces are multi-valued functions
of the system parameters.

Exceptional points for a particular system occur when the roots of the system,
defined by the equation $|\mathcal{D}(\omega)|=0$ 
also satisfy $\frac{d}{d \omega}|\mathcal{D}(\omega)|=0$,
\citep{Heiss:2004b} where ${\mathcal{D}}$
is an $n \times n$ matrix defined by the system's equations of motion.
In the case of a Vlasov-Maxwell system $|\mathcal{D}(\omega)|$ is the dispersion relation 
derived from the homogeneous wave equation, given by Equation 10-73 in 
Stix 1992.\citep{Stix:1992}
As the Vlasov-Maxwell system of equations is complicated, we leave 
numerical or analytical treatments for the locations of exceptional points
to a later paper.
The multivalued nature of the dispersion surface is not simply a 
mathematical abstraction, but
has been shown to be a physical reality in several laboratory
experiments,\citep{Heiss:2004a,Heiss:2004b}
and has been identified in several plasma descriptions, including
gyrokinetics.\citep{Kammerer:2008} In this appendix
we identify two examples of exceptional points in the Vlasov-Maxwell system,
one at large MHD scales and a second at parallel wavelengths near
the proton inertial length.

\begin{figure*}[h]
\begin{center}
\includegraphics[width=18.cm,viewport=15 0 495 150, clip=true]
{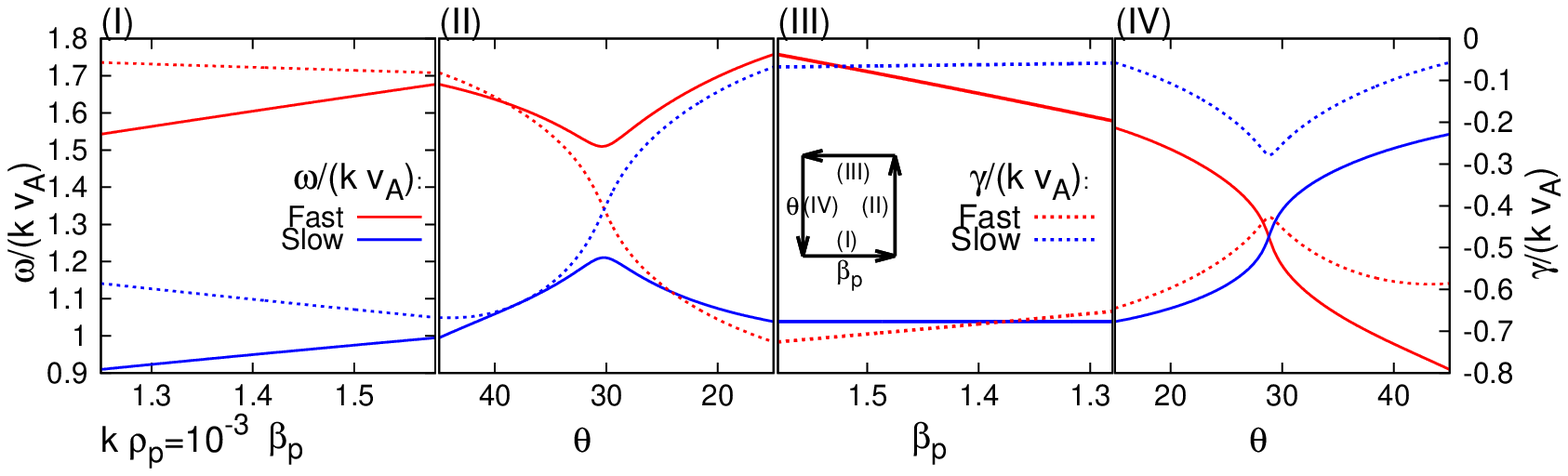}\\
\includegraphics[width=5.5cm,viewport=15 0 150 150, clip=true]
{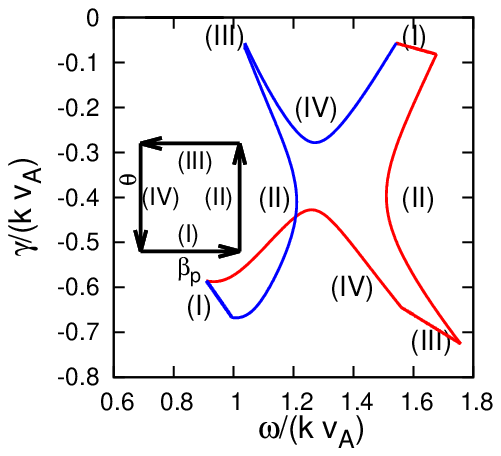}
\includegraphics[width=5.5cm,viewport=15 0 150 150, clip=true]
{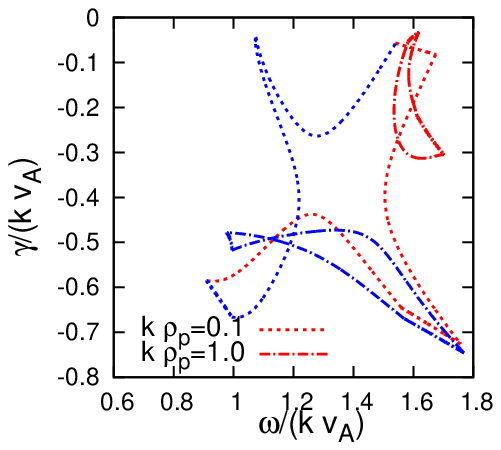}\\
\caption{
The transition between slow (blue lines) and fast (red)
modes due to a large-scale exceptional point.
The frequency $\omega/k v_A$ and damping rate $\gamma/k v_A$ for the two modes 
as a function of $\beta_p$ and $\theta$ along the path prescribed in the text 
are plotted in the top row.
The parametrized path for these two modes in complex frequency space is given
in the lower left panel for $k \rho_p=10^{-3}$, and in the lower right
panel for $k \rho_p=0.1$ and $1.0$.
}
\label{fig:EP}
\end{center}
\end{figure*}

The interchange of large scale fast and slow
Vlasov-Maxwell waves,\citep{Gary:1993,Krauss-Varban:1994,Klein:2012} analogous to the
interchange between the MHD fast and slow modes, is due to an exceptional point 
at $(\beta_p, \theta) \approx (1.3,30^\circ)$, where $\theta$ is the angle between
$\V{k}$ and $\V{B}_0$. This exceptional point, illustrated in the three
panels of Figure~\ref{fig:EP}, exists for large wavelengths with $k \rho_p < 1.0$.
To encircle the exceptional point, 
we follow the complex frequency solution for the fast and slow modes
along four paths in $(\beta_p,\theta)$ space: from 
\textbf{(I)} $(1.25,45^\circ)$ to $(1.58,45^\circ)$,
\textbf{(II)} $(1.58,45^\circ)$ to $(1.58,15^\circ)$,
\textbf{(III)} $(1.55,15^\circ)$ to $(1.25,15^\circ)$, and
\textbf{(IV)} $(1.25,15^\circ)$ back to $(1.25,45^\circ)$.
All other plasma parameters for the system are fixed: $k \rho_p=10^{-3}$, 
$T_\perp/T_\parallel|_p=T_\perp/T_\parallel|_e=1.0$, $T_p/T_e=1.0$, and $v_{tp}/c=10^{-4}$.
The frequency and damping rates for the two modes along these four
paths are presented as functions of either $\beta_p$ or $\theta$ in 
the top row of Figure~\ref{fig:EP}. 
We see that after following the prescribed path the two modes have exchanged 
complex frequency solutions. 
The continuous nature of this
mode exchange is shown in the lower left panel of Figure~\ref{fig:EP},
in which the same solutions for the fast and slow modes are plotted
in complex frequency space over paths \textbf{I}-\textbf{IV}. 
This exceptional point exists for wave lengths up to $k \rho_p \simeq 1$, as
is shown in the bottom right panel, where the linear solutions 
are calculated using the same paths in $(\beta_p,\theta)$ with 
$k \rho_p=0.1$ (dashed lines) and $1.0$ (dot-dash). For the latter,
smaller scale case, the exceptional point has disappeared
and the fast and slow modes are no longer connected
by a continuous variation of $\beta_p$ and $\theta$.

\begin{figure}[h]
\begin{center}
\includegraphics[width=12.5cm,viewport=5 5 200 200, clip=true]
{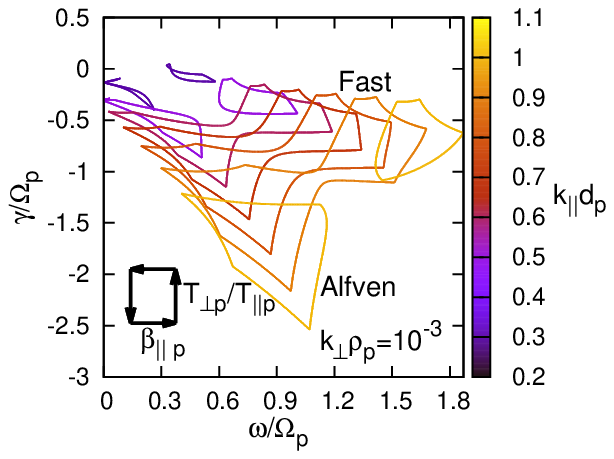}
\caption{
Parametrized solutions in complex frequency space of the \Alfven and fast modes 
along paths of \Bpar $\in [7.0,10.0]$ and \tempAni $ \in [0.7,1.0]$,
with the value of $k_\parallel d_p \in [0.2,1.0]$ given by the color bar.
The mode conversion between the fast and \Alfven modes for  
$k_\parallel d_p \in [0.6,0.9]$ is indicative of the presence of 
a small-scale exceptional point as described in the text.
}
\label{fig:EP_kpar}
\end{center}
\end{figure}

While the large scale $(\beta_p$, $\theta)$ exceptional point
disappears at small wavelengths, a variety of other exceptional points arise near
$k_\parallel d_p \simeq 1$ and $k_\perp \rho_p \simeq 1$ connecting two, or more, modes
through the variation of other plasma parameters. An example of
such a small scale exceptional point is found near $\beta_{\parallel p} \approx 9.0$
and $T_\perp/T_\parallel|_p \approx 0.9$. This exceptional point, 
illustrated in Figure~\ref{fig:EP_kpar}, connects the \Alfven and 
fast solutions. 
The values for the fast and \Alfven solutions
in complex frequency space along the paths connecting 
$(\beta_{\parallel p},T_{\perp p}/T_{\parallel p})=(7.0,1.0), \ (7.0,0.7), \ (10.0,0.7),$ and $(10.0,1.0)$
are calculated for a set of $k_\parallel d_p \in [0.2,1.0]$.
All other plasma parameters are fixed at
$k_\perp \rho_p = 10^{-3}$, $T_{\parallel p}/T_{\parallel e}=1$,
$T_{\perp e}/T_{\parallel e}=1$ and $v_{t \parallel p}/c = 10^{-4}$.
The value of $k_\parallel d_p$ is indicated in Figure~\ref{fig:EP_kpar}
in color.
Unlike the large scale $(\beta_p,\theta)$ exceptional point,
which persists over several orders of magnitude in wavelength $k$, this
fast-\Alfven exceptional point is narrowly restricted to 
$k_\parallel d_p \in [0.6,0.9]$; for scales above and below this range,
the \Alfven and fast solutions remain distinct. The effects of
this exceptional point can be seen in the left column of
Figure~\ref{fig:aleph_scan}, where the $\beta_{\parallel p}=10.0$ and 
\tempAni $<1.0$ \Alfven solution with $k_\parallel d_p = 0.6$ becomes unstable
to the parallel firehose instability, the instability associated with
the fast/whistler wave, a behavior not seen for any of the $\beta_p < 10$
\Alfven solutions. Note that the $\beta_p=10.0$ fast solution 
with $k_\parallel d_p = 0.5$,
shown in the second column of Figure~\ref{fig:aleph_scan},
does not exhibit this interchange of solutions, illustrating the
strong dependence of this exceptional point on $k_\parallel d_p$.

Proper identification of linear solutions is complicated by the 
existence of exceptional points for Vlasov-Maxwell dispersion surfaces. 
The existence of these points
allows for a continuous change in frequency and eigenfunction characteristics from
one linear solution to another, allowing for possibility of coupling between
distinct linear solutions and complicating any attempt to understand the nature
of turbulent energy transfer and dissipation in terms of the nature of the underlying
linear fluctuations. Understanding the behavior of these exceptional points
for collisionless plasmas and constraining their impact on the solar wind will be left to
later work.


%



\end{document}